\newcommand\apjcls{1}
\newcommand\aastexcls{2}
\newcommand\othercls{3}

\newcommand\papercls{\aastexcls}
\documentclass[tighten, times, preprint2]{aastex6}  

\if\papercls \apjcls
\usepackage{apjfonts}
\else\if\papercls \othercls
\usepackage{epsfig}
\fi\fi
\usepackage{ifthen}
\usepackage{natbib}
\usepackage{amssymb, amsmath}
\usepackage{appendix}
\usepackage{etoolbox}
\usepackage[T1]{fontenc}
\usepackage{paralist}

\if\papercls \apjcls
\newcommand\aas{\ref@jnl{AAS Meeting Abstracts}}
\newcommand\dps{\ref@jnl{AAS/DPS Meeting Abstracts}}
\newcommand\maps{\ref@jnl{MAPS}}
\else\if\papercls \othercls
\usepackage{astjnlabbrev-jh}
\fi\fi

\if\papercls \aastexcls
\hypersetup{citecolor=blue, 
            linkcolor=blue, 
            menucolor=blue, 
            urlcolor=blue}  
\else
\usepackage[
bookmarks=true,           
bookmarksnumbered=true,   
colorlinks=true,          
citecolor=blue,           
linkcolor=blue,           
menucolor=blue,           
urlcolor=blue,            
linkbordercolor={0 0 1},  
pdfborder={0 0 1},
frenchlinks=true]{hyperref}
\fi

\if\papercls \othercls

\else

\fi

\providecommand{\adsurl}[1]{\href{#1}{ADS}}

\makeatletter
\patchcmd{\NAT@citex}
  {\@citea\NAT@hyper@{%
     \NAT@nmfmt{\NAT@nm}%
     \hyper@natlinkbreak{\NAT@aysep\NAT@spacechar}{\@citeb\@extra@b@citeb}%
     \NAT@date}}
  {\@citea\NAT@nmfmt{\NAT@nm}%
   \NAT@aysep\NAT@spacechar\NAT@hyper@{\NAT@date}}{}{}

\patchcmd{\NAT@citex}
  {\@citea\NAT@hyper@{%
     \NAT@nmfmt{\NAT@nm}%
     \hyper@natlinkbreak{\NAT@spacechar\NAT@@open\if*#1*\else#1\NAT@spacechar\fi}%
       {\@citeb\@extra@b@citeb}%
     \NAT@date}}
  {\@citea\NAT@nmfmt{\NAT@nm}%
   \NAT@spacechar\NAT@@open\if*#1*\else#1\NAT@spacechar\fi\NAT@hyper@{\NAT@date}}
  {}{}
\makeatother

\makeatletter
\DeclareRobustCommand{\lowcase}[1]{\@lowcase#1\@nil}
\def\@lowcase#1\@nil{\if\relax#1\relax\else\MakeLowercase{#1}\fi}
\makeatother

\DeclareSymbolFont{UPM}{U}{eur}{m}{n}
\DeclareMathSymbol{\umu}{0}{UPM}{"16}
\let\oldumu=\umu
\renewcommand\umu{\ifmmode\oldumu\else\math{\oldumu}\fi}

\if\papercls \othercls

\else

\fi

\let\oldsim=\sim
\renewcommand\sim{\ifmmode\oldsim\else\math{\oldsim}\fi}
\let\oldpm=\pm
\renewcommand\pm{\ifmmode\oldpm\else\math{\oldpm}\fi}
\newcommand\by{\ifmmode\times\else\math{\times}\fi}

\newcommand\tablebox[1]{\begin{tabular}[t]{@{}l@{}}#1\end{tabular}}
\newbox{\wdbox}
\renewcommand\c{\setbox\wdbox=\hbox{,}\hspace{\wd\wdbox}}
\renewcommand\i{\setbox\wdbox=\hbox{i}\hspace{\wd\wdbox}}

\newcount\timect
\newcount\hourct
\newcount\minct
\newcommand\now{\timect=\time \divide\timect by 60
         \hourct=\timect \multiply\hourct by 60
         \minct=\time \advance\minct by -\hourct
         \number\timect:\ifnum \minct < 10 0\fi\number\minct}

\catcode`@=11

\newcommand\comment[1]{}

\newcommand\commenton{\catcode`\%=14}

\renewcommand\math[1]{$#1$}
\newcommand\mathshifton{\catcode`\$=3}

\let\atab=&
\newcommand\atabon{\catcode`\&=4}

\let\oldmsp=\sp
\let\oldmsb=\sb
\def\sp#1{\ifmmode
           \oldmsp{#1}%
         \else\strut\raise.85ex\hbox{\scriptsize #1}\fi}
\def\sb#1{\ifmmode
           \oldmsb{#1}%
         \else\strut\raise-.54ex\hbox{\scriptsize #1}\fi}
\newbox\@sp
\newbox\@sb
\def\sbp#1#2{\ifmmode%
           \oldmsb{#1}\oldmsp{#2}%
         \else
           \setbox\@sb=\hbox{\sb{#1}}%
           \setbox\@sp=\hbox{\sp{#2}}%
           \rlap{\copy\@sb}\copy\@sp
           \ifdim \wd\@sb >\wd\@sp
             \hskip -\wd\@sp \hskip \wd\@sb
           \fi
        \fi}
\def\msp#1{\ifmmode
           \oldmsp{#1}
         \else \math{\oldmsp{#1}}\fi}
\def\msb#1{\ifmmode
           \oldmsb{#1}
         \else \math{\oldmsb{#1}}\fi}

\def\supon{\catcode`\^=7}

\def\subon{\catcode`\_=8}

\def\supsubon{\supon \subon}

\newcommand\actcharon{\catcode`\~=13}

\newcommand\paramon{\catcode`\#=6}

\comment{And now to turn us totally on and off...}

\newcommand\reservedcharson{ \commenton  \mathshifton  \atabon  \supsubon 
                             \actcharon  \paramon}

\catcode`@=12
\reservedcharson

\if\papercls \apjcls

\else

\fi

\if\papercls \othercls
\else
  \newcommand\inpress{n}
  \if\inpress y
    \received{\today}
    \revised{}
    \accepted{}
    \if\papercls \apjcls
    \slugcomment{}
    \fi
  \else
  \slugcomment{\tablebox{In preparation for {\em ApJ}. DRAFT of {\today}.}}
  \fi
\fi

\newcommand\chisq{\ifmmode{\chi\sp{2}}\else\math{\chi\sp{2}}\fi}
\newcommand\redchisq{\ifmmode{ \chi\sp{2}\sb{\rm red}}
                    \else\math{\chi\sp{2}\sb{\rm red}}\fi}
\newcommand\Teq{\ifmmode{T\sb{\rm eq}}\else$T$\sb{eq}\fi}
\newcommand\mjup{\ifmmode{M\sb{\rm Jup}}\else$M$\sb{Jup}\fi}
\newcommand\rjup{\ifmmode{R\sb{\rm Jup}}\else$R$\sb{Jup}\fi}
\newcommand\msun{\ifmmode{M\sb{\odot}}\else$M\sb{\odot}$\fi}
\newcommand\rsun{\ifmmode{R\sb{\odot}}\else$R\sb{\odot}$\fi}
\newcommand\mearth{\ifmmode{M\sb{\oplus}}\else$M\sb{\oplus}$\fi}
\newcommand\rearth{\ifmmode{R\sb{\oplus}}\else$R\sb{\oplus}$\fi}

\shorttitle{Infall of NGC 1404}
\shortauthors{Sheardown et al.}

\begin{document}

\title{The Recent Growth History of the Fornax Cluster Derived from Simultaneous Sloshing and Gas Stripping: Simulating the Infall of NGC 1404}

\author{Alex~Sheardown\altaffilmark{1}, Elke~Roediger\altaffilmark{1}, Yuanyuan~Su\altaffilmark{2}, Ralph~P.~Kraft\altaffilmark{2}, Thomas~Fish\altaffilmark{1}, John~A.~ZuHone\altaffilmark{2}, William~R.~Forman\altaffilmark{2}, Christine~Jones\altaffilmark{2}, Eugene~Churazov\altaffilmark{3}, Paul~E.~J.~Nulsen\altaffilmark{2}}
\affil{\sp{1} E.A. Milne Centre for Astrophysics, School of Mathematics and Physical Sciences, University of Hull, Hull, HU6 7RX, United Kingdom} 
\affil{\sp{2} Harvard-Smithsonian Center for Astrophysics, 60 Garden Street, Cambridge, MA 02138, USA}
\affil{\sp{3}Max Planck Institute for Astrophysics, Karl-Schwarzschild-Str. 1, 85741, Garching, Germany}

\email{A.Sheardown@2011.hull.ac.uk}

\begin{abstract}
We derive the recent growth history of the Fornax Cluster, in particular the recent infall of the giant elliptical galaxy NGC 1404. We show, using a simple cluster minor merger simulation tailored to Fornax and NGC 1404, that a second or more likely third encounter between the two reproduces all main merger features observed in both objects; we firmly exclude a first infall scenario. Our simulations reveal a consistent picture: NGC 1404 passed by NGC 1399 about 1.1 - 1.3 Gyrs ago from the NE to the SW and is now almost at the point of its next encounter from the S. This scenario explains the sloshing patterns observed in Fornax - a prominent northern cold front and an inner southern cold front. This scenario also explains the truncated atmosphere, the gas stripping radius of NGC 1404, and its faint gas tail. Independent of the exact history, we can make a number of predictions. A detached bow shock south of NGC 1404 should exist which is a remnant of the galaxy's previous infall at a distance from NGC 1404 between 450 - 750 kpc with an estimated Mach number between 1.3 and 1.5. The wake of NGC 1404 also lies S of the galaxy with enhanced turbulence and a slight enhancement in metallicity compared to the undisturbed regions of the cluster. SW of NGC 1404, there is likely evidence of old turbulence originating from the previous infall. No scenario predicts enhanced turbulence outside of the cold front north west of the cluster center.

\end{abstract}


\section{INTRODUCTION}
\label{sec:introduction}

Embedded in the large scale structure, galaxy clusters are the largest gravitationally bound systems in the universe containing hundreds or more galaxies. Under the framework of the hierarchical model, clusters are still growing through sequential mergers and accretion of smaller systems - from subclusters to galaxy groups to the infall of galaxies. Studying the dynamics of cluster mergers is particularly suited to the X-ray regime, as gas rich mergers leave a clear trace of the merger history due to the thermal bremsstrahlung emission of the intra-cluster medium (ICM). Mergers have a significant impact on the thermal state of the cluster by inducing bulk motions, driving shocks in the ICM, and generating regions of turbulence which then dissipate and heat the surrounding gas (\citealp{Roediger2009b}, \citealp{Bykov2015}). Excellent examples of such merger shocks can be seen in the galaxy cluster 1E 0657-56 commonly known as the Bullet Cluster \citep{Markevitch2002a} and in Abell 520 \citep{Markevitch2004}. Further merger shocks have also been observed in Abell 85 \citep{Ichinohe2015a}, Abell 2146 \citep{Russell2010a}, and Abell 665 \citep{Dasadia2016}. \par

Using mass ratios, mergers can roughly be distinguished into two main regimes - major mergers and minor mergers. The former occur between approximately equal mass systems such as two clusters (a 1:1 mass ratio), whereas minor mergers involve a low and a high mass system, where the infall of an early-type galaxy into a cluster can be regarded as a very minor merger. In what follows below, for clarity we will refer to the lower mass merger partner as "the galaxy", however all explanations are valid for subclusters as well. \par

\begin{figure*}
\centering
\includegraphics[scale=0.54]{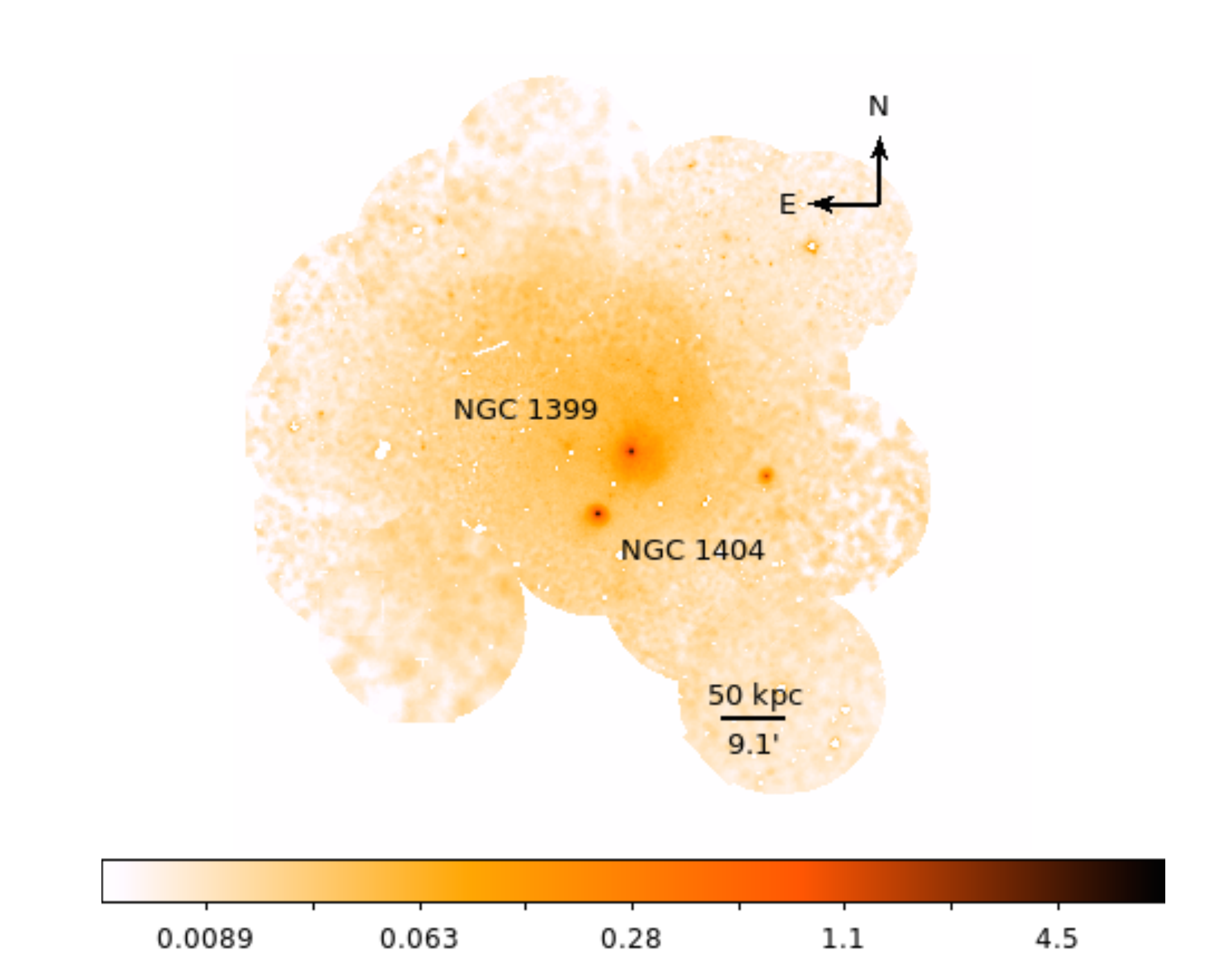}
\includegraphics[scale=0.54]{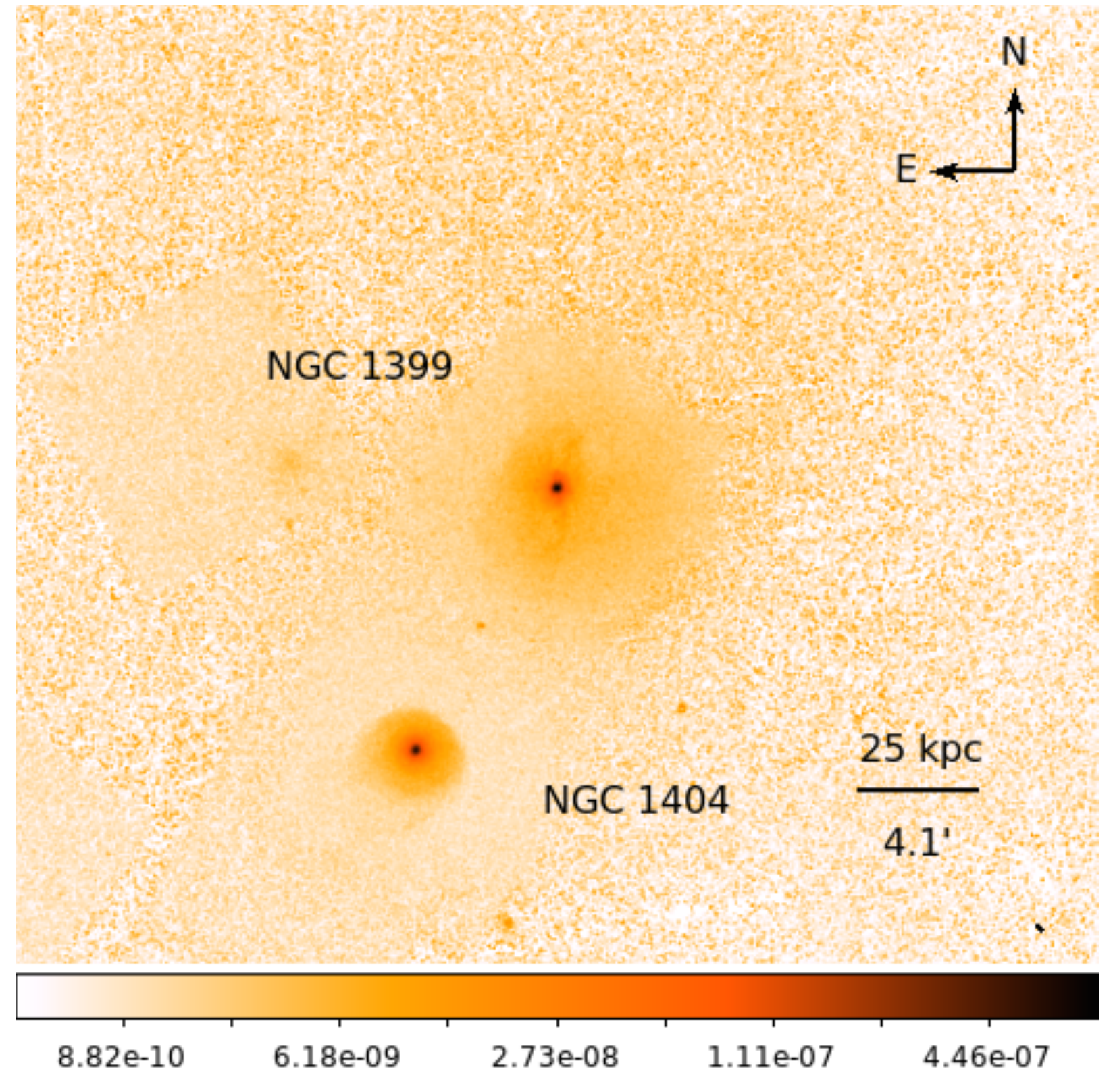}
\caption{Top: XMM image of Fornax in the energy band of 0.5 - 2.0 keV taken from \citet{Su2017d} in units of photons/s/cm$^{2}$/deg$^{2}$. A prominent cold front is evident to the north of the cluster center as well as a smaller cold front to the south due to sloshing in the cluster gas. Bottom: Taken from \citet{Su2017f}, exposure-corrected with blank-sky background subtracted Chandra mosaic image of Fornax in the energy band of 0.5 - 2.0 keV and in the unit of photon cm$^{-2}$s$^{-1}$. Chandra gives a clear view of the truncated atmosphere of NGC 1404 as well as its faint gas tail indicative of a galaxy being gas stripped due to infall.}
\label{fig:obsxmm}
\end{figure*} 

Consider the scenario of a small subcluster merging with a larger main cluster. Two processes occur simultaneously: the sloshing of the main (larger) cluster and gas stripping of the infalling galaxy or subcluster. Gas sloshing in the main cluster occurs as the subcluster moves through the pericenter, offsetting the ICM and dark matter in the main cluster core which then begins to move towards the subcluster as it is pulled gravitationally towards it. During this phase, ram pressure can act to decelerate the ICM gas separating it from the dark matter as the dark matter is still free to move towards the subcluster. As the subcluster completes its passage, the offset ICM gas falls back towards the main cluster core generating cold fronts that propagate outwards (\citealp{Ascasibar2006}, \citealp{ZuHone2009a}, Figures 12 and 14 in \citealp{Bykov2015}). The outward propagation of the cold fronts depends mainly on the cluster potential and ICM profiles. Therefore the recent merger history of a given cluster can be reconstructed from the observed cold fronts. The sloshing cold fronts are sharp contact discontinuities in the ICM density and temperature (but not pressure) that can be seen through X-ray observations. Although cold fronts appear to be stable, velocity shears are likely present giving rise to Kelvin-Helmholtz instabilities (KHI) \citep{Roediger2013c}. However, the strength of the ICM viscosity and magnetic field can act to potentially suppress the instability (\citealp{Vikhlinin2001}, \citealp{ZuHone2011a}). \par

The second process is the gas stripping of the infalling galaxy. As the galaxy moves through the ICM, it is progressively stripped due to a ram pressure which causes a drag force on the galaxy which in turn strips away its gaseous atmosphere (\citealp{Gunn1972}, \citealp{Larson1980a}, \citealp{Roediger2008b}, \citealp{Roediger2015c}, \citealp{DeGrandi2016b}). Additionally, the galaxy can be stripped via Kelvin-Helmholtz instabilities which arise due to velocity shears between the ICM gas and the galactic atmosphere. As the ram pressure is P$_\text{ram}$ = $\rho_\text{ICM}$v$_\text{gal}^{2}$, the strength of the gas stripping is dependent on the galaxies orbit through the cluster, in particular the pericenter distance to the cluster center. This determines the density of the ICM that the galaxy will experience and the orbital velocity of the galaxy. When the ram pressure is great enough, the stripped gas appears as an X-ray tail as observed in several elliptical galaxies; NGC 4552 \citep{Machacek2006a}, NGC 4406 \citep{Randall2008a}, NGC 4472 \citep{Kraft2011a}, NGC 1400 \citep{Su2014b}, CGCG254-021 in Zwicky 8338 \citep{Schellenberger2015a}. Thus, an infalling galaxy can be characterised by a leading upstream edge which hosts a truncated atmosphere along with a downstream tail of stripped galactic gas. In the tail, the stripped gas should mix with the ambient ICM, unless mixing is suppressed by e.g. viscosity or magnetic fields. The exact state of the tail gas can thus be potentially used to determine the transport properties of the ICM (\citealp{Roediger2015d}, \citealp{Su2017d}). \par
 
Using our tailored hydro+Nbody simulations, we present a case study of a merger between an infalling galaxy and a cluster, that of the elliptical galaxy NGC 1404 and the Fornax Cluster. The motivation in choosing to model this system is due to its relatively simple components which are not complicated by galaxy-galaxy interactions or AGN outbursts, coupled with its relatively close proximity which offers extensive observational data in the X-ray and optical regimes. In this regard, NGC 1404 offers a unique probe to study transport processes along with cluster wide physics due to its pre-truncated atmosphere. Thus, the uncertainty regarding its initial gas contents and spatial configuration is unimportant. Our aim is to simulate a simple cluster minor merger between NGC 1404 and Fornax by using appropriate gravitational potentials and gas contents which agree with observationally derived profiles. Comparing the resulting sloshing and stripping features to the real observations, we determine the recent merger history of Fornax. In Section \ref{sec:ngcandf}, we describe the target galaxy and cluster, detailing key features of both systems. In Section \ref{sec:method}, we outline the initial conditions used for generating our model, designed based on observational constraints. Section \ref{sec:results} presents the result of the simulation while section \ref{sec:discussion} discusses implications for the history and physics of the Fornax cluster. In section \ref{sec:summary}, we summarise our findings.

\section{Setting the Scene: NGC 1404 and the Fornax Cluster}
\label{sec:ngcandf}

The Fornax Cluster is a nearby, low mass, cool core galaxy cluster located in the southern hemisphere at a distance of 19 Mpc ($1'$ = 5.49 kpc) and a redshift of z = 0.00475 \citep{Paolillo2002a}. Due to its close proximity, Fornax has been extensively imaged in a number of wavelengths by a range of telescopes and instruments, in particular in X-rays by ROSAT, Chandra and XMM-Newton, with the latter two having the ability to resolve structures in Fornax down to 100 pc. Schematically speaking, the main body of Fornax is dominated by the brightest cluster galaxy NGC 1399, a large almost spherical elliptical galaxy (E1), with another sub system situated > 1 Mpc south west of NGC 1399 centered around Fornax A. The Fornax core, centered around NGC 1399, is encapsulated by the ICM with a temperature of $\sim$ 1.5 keV (\citealp{Rangarajan1995a}, \citealp{Jones1997a}, \citealp{Paolillo2002a}, \citealp{Machacek2005a}, \citealp{Su2017f}). In terms of size, using a scaling relation based on the ICM temperature, \citet{Su2017f} estimated the virial mass of Fornax to be r$_{vir}$ $\approx$ 750 kpc and \citet{Drinkwater2001a} calculated a dynamical mass of 7 $\pm$ 2 $\times$10$^{13}$ M${_\odot}$ within a projected radius of 1.4 Mpc. Using joint Chandra and XMM observations of Fornax, \citet{Su2017d} revealed evidence of asymmetry and merger induced gas sloshing occurring in the cluster core, in particular identifying four sloshing cold fronts as was suggested by \citet{Su2017f}. From joint Suzaku and XMM observations, \citet{Murakami2011c} also found evidence of asymmetry by analysing temperature and metallicity distributions in Fornax. They found that the region $13'$ (71 kpc) north of the cluster center has a low ICM temperature and high Fe abundance in comparison to the region \sim $17' - 27'$ (93 kpc - 148 kpc) south of the cluster center, pertaining to recent dynamical evolution. \par

NGC 1404 is an elliptical galaxy and the second brightest galaxy in Fornax, located south east to the central galaxy, NGC 1399, at a projected radius of \sim 60 kpc. The galaxy's atmosphere harbours a sharp upstream leading edge 8 kpc from its center forming a cold front towards NGC 1399 and an \sim 8 kpc long gaseous tail to the south east (\citealp{Jones1997a}, \citealp{Machacek2005a}, \citealp{Su2017f}). Using stagnation point analysis, \citet{Machacek2005a} determined the galaxy to be in the same plane of the sky as the cluster center and provided an estimate for the Mach number of NGC 1404 to be 0.83-1.03 with a relative velocity to the ICM being 531-657 km s$^{-1}$, whilst \citet{Scharf2005a} estimate a Mach number of 1.3 $\pm$ 0.3 and a velocity of 660 $\pm$ 260  km s$^{-1}$. Using a 670 ks Chandra observation, \citet{Su2017f} led an extensive study based on stagnation point pressure analysis to determine that the galaxy is infalling at an inclination angle of 33$^{\circ}$ with a Mach number of 1.32. Inside the leading edge of the galaxy, they calculate an electron density of n$_{e}$ = 6.1 $\times$ 10$^{-3}$ cm$^{-3}$ with a gas temperature of 0.6 \pm 0.02 keV. Further, they calculate that the tail of NGC 1404 is 16 kpc wide and 8 kpc in length in projection with a gas temperature in the region of 0.9-1.0 keV. They suggest that since the temperature in the gas tail is consistently hotter than the remnant core of the galaxy (0.6 keV) and cooler than the ambient ICM gas (1.5 keV), that thermal conduction is heating the stripped gas and/or turbulent mixing of the ICM gas is happening downstream in the tail. \par

Another indication for an ongoing merger between NGC 1404 and Fornax is the globular cluster content of both systems. In particular, \citet{Forbes1998a} and \citet{Bekki2003a} both find that NGC 1399 has rich globular cluster content (high specific frequency) whilst NGC 1404 has poor globular cluster content (low specific frequency) compared to the average for elliptical galaxies, suggesting that NGC 1399 may have stripped NGC 1404 of some its globular clusters as it undergoes a merger. Under this argument it would imply that NGC 1404 has already fallen through the cluster once already.

\section{Simulation Setup}
\label{sec:method}

\subsection{Initial Conditions}
To tailor our simulations to NGC 1404 and Fornax, we aimed to match their gravitational potentials and gas atmospheres to observations. This information is available through observations of stellar light, stellar velocity dispersion, ICM temperature, pressure, and density distribution from X-rays. We aim to match the observational data at the end of the simulation run, this required some test runs to find initial conditions such that the evolved cluster, after the merger, matches the data. Our models for NGC 1404 and Fornax are set up to be spherically symmetric, self gravitating and in hydrostatic equilibrium following the set up procedure as described in \citet{ZuHone2011a}. Each simulation has a different merger time and geometry (explained in Section \ref{sec:orb}). Therefore, to get a perfect match for each would require a slightly different initial model for each simulation. This wouldn't only be impractical, but would also prevent an easy comparison between the different simulations. Thus, the initial model we have chosen is a suitable compromise for making our simulations representative of NGC 1404 and Fornax, while being practical at the same time. Furthermore, we experimented with a range of initial models similar to the one presented here, and the main conclusion is independent of the exact choice.

\subsubsection{Fornax}
\begin{figure}[h!]
\centering
\includegraphics[scale=0.24]{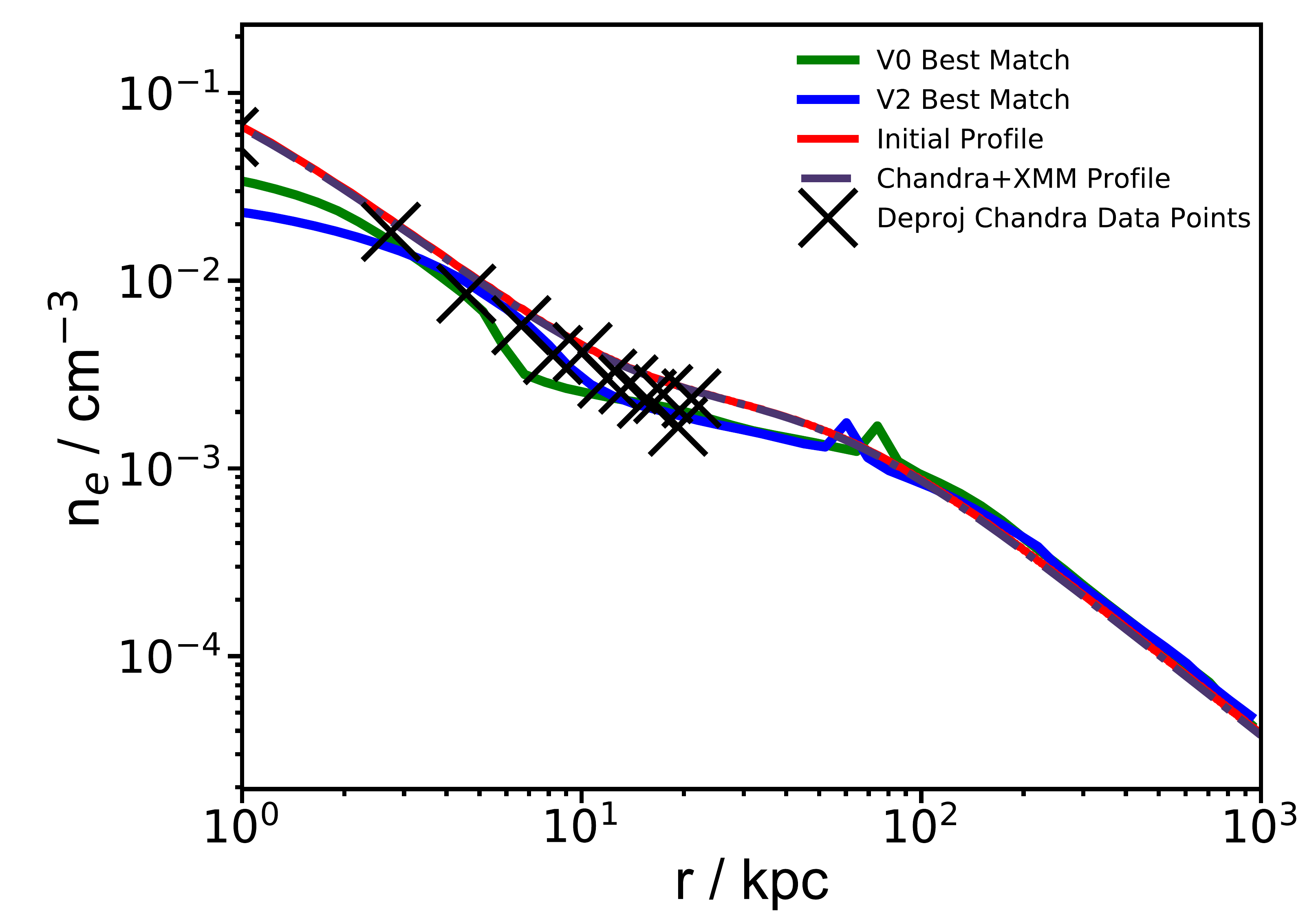}
\includegraphics[scale=0.24]{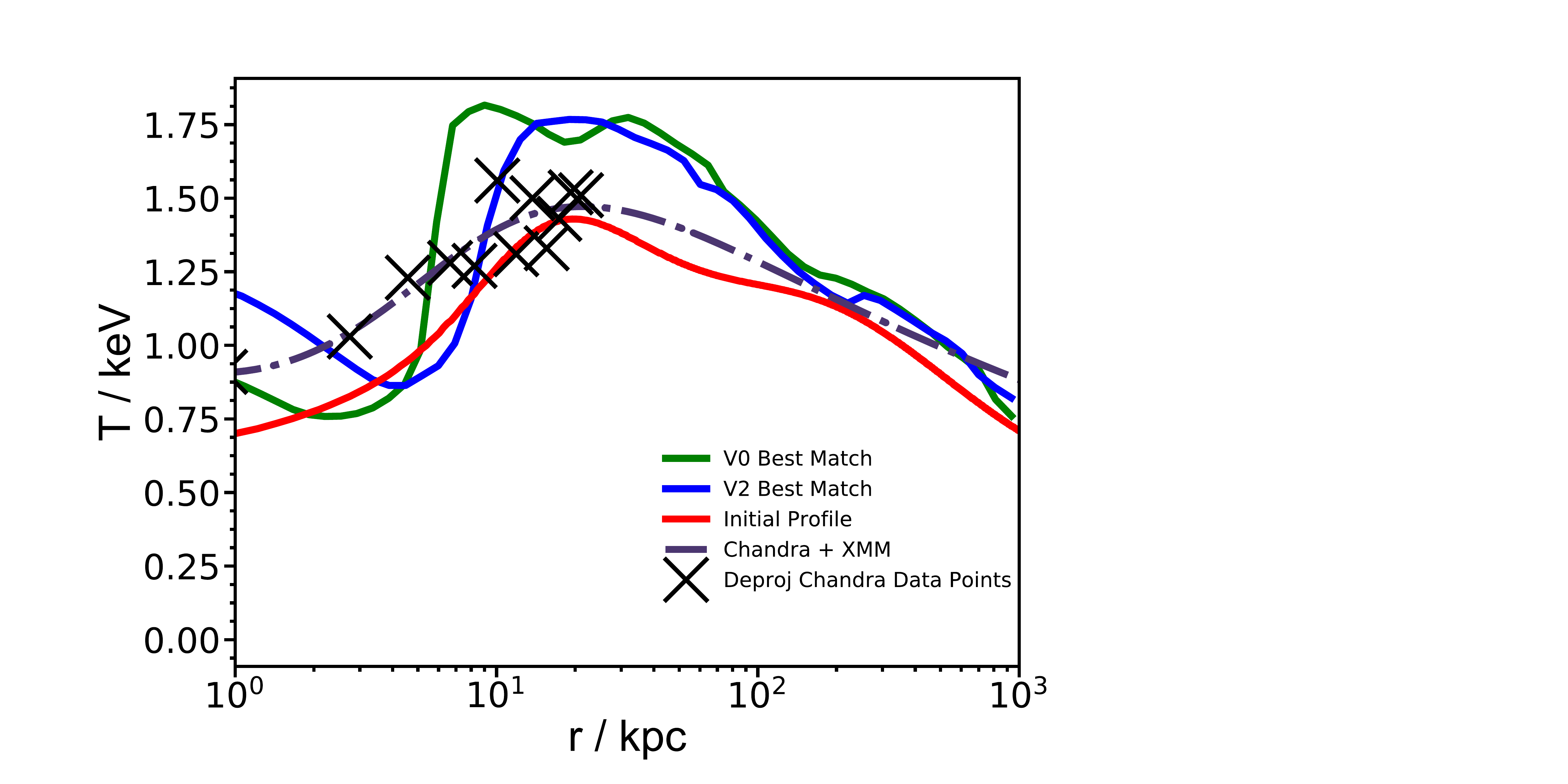}
\caption{Comparison of simulated and observed ICM profiles for the Fornax cluster from the V0 and V2 simulations. Observed profiles are spherically averaged. Profiles from the simulation are taken at the final merger state which provides a visual match to observation and are spherically averaged. Top: Electron density profiles. Bottom: Temperature profiles. The overall ICM distribution in the observed and simulated Fornax cluster agree well, considering our simple model. The decrease in density at the very centre of Fornax in our profiles is partly due to gas stripping and partly due to resolution.}
\label{fig:fornaxprofiles}
\end{figure}

To tailor our simulation to Fornax, we use Chandra and XMM-Newton data to model the Fornax ICM gas density in the form of a double $\beta$ profile. Chandra data covers the inner 25 kpc of Fornax while the XMM data reaches out to 200 kpc; we extrapolate the observational results out to larger radii. Figure \ref{fig:fornaxprofiles} compares the Fornax ICM density and temperature profiles taken from Chandra and XMM with our model profiles for the V0 and V2 simulations. The parameters for the double beta model are presented in Table \ref{table:fparameters}. The match and deviations between observations and model are explained below. \par

\begin{figure}[h!]
\centering

\includegraphics[scale=0.42]{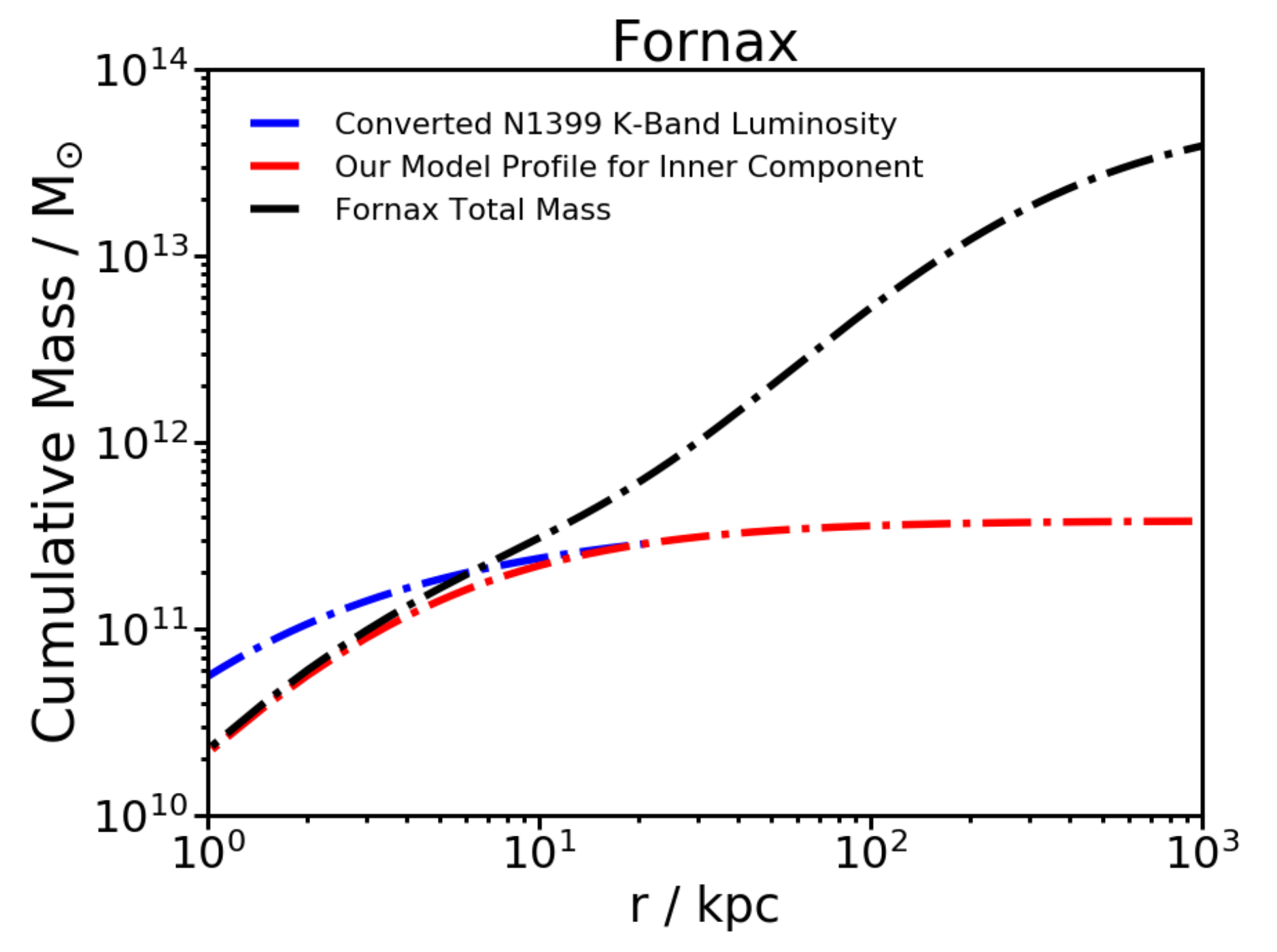}
\caption{Cumulative mass profile for our model Fornax cluster compared to cumulative stellar mass of NGC 1399, converted from K-Band luminosity. The total mass of the model cluster consists of an inner and outer Hernquist potential, where roughly the inner potential represents the NGC 1399 stellar component. As a compromise between the stellar light data and the observed ICM temperature profile, we chose an inner component somewhat less compact than observed because otherwise the hydrostatic initial setup of the cluster results in an unrealistically high central ICM temperature.}
\label{fig:fcmass}
\end{figure}

We model the total gravitational potential of Fornax with a double Hernquist potential (eq. 1): 
\begin{equation} \label{eq}
\begin{split}
 \mathrm{\rho(r)}  & = \mathrm{ \frac{M_{\text{dm}}}{2\pi a_{\text{dm}}^{3}} \frac{1}{\frac{r}{a_{\text{dm}}}(1+\frac{r}{a_{\text{dm}}})^{3}}  + \frac{M_{*}}{2\pi a_{*}^{3}} \frac{1}{\frac{r}{a_{*}}(1+\frac{r}{a_{*}})^{3}} }\\ 
 &  \mathrm {M(< r) = M_{dm} \frac{r^{2}}{(r+a_{dm})^{2}} } +  M_{*} \frac{r^{2}}{(r+a_{i*)^{2}} }
 \end{split}
 \end{equation} 
 
where M$_{\text{dm}}$ and a$_{\text{dm}}$ are the mass and scale length for the outer component of Fornax, and M$_{*}$ and a$_{*}$ are the mass and scale length for the inner component respectively. This model is chosen for the useful property of a finite total mass and thus does not require truncation like an NFW profile. Further, using two components allows us to better capture the inner and outer potentials of Fornax. Although we do not distinguish between luminous and dark matter, we can think of one potential (the outer component) as the dark matter content of Fornax described by mass M$_{\text{dm}}$ and scale length a$_{\text{dm}}$, and the inner component as the dominant central galaxy in Fornax, NGC 1399 described by mass M$_{*}$ and scale length a$_{*}$. We note that this total potential includes the ICM as well. The particle density in the simulation is set as the difference between the total density and the ICM density. For the "dark matter" component of the Hernquist potential, we have a constraint for the total mass of Fornax from \citet{Drinkwater2001a} of 7 $\pm$ 2 $\times$10$^{13}$ M${_\odot}$ based on the method of \citet{Diaferio1999a}. Furthermore, M$_{\text{dm}}$ and a$_{\text{dm}}$ affect the overall ICM temperature profile, which is constrained by the X-ray data. Thus we select a dark matter mass of 6 $\times$ 10$^{13}$ M$_{\odot}$ with a scale length of 250 kpc for the parameters M$_{\text{dm}}$ and a$_{\text{dm}}$. For the inner potential we are guided by the stellar light. We convert the K-band luminosity profile of NGC 1399 to a cumulative mass profile using a stellar mass-to-light ratio in the K-band of 1.3 $\frac{M_{\odot}}{L_{\odot}}$ (taken from \citealp{Silva1998a}). However, the central potential component also impacts the central ICM temperature profile. A very steep central potential that closely matches the stellar light data leads to unrealistically high central ICM temperatures. Therefore, we find that a stellar mass of M$_{*}$ = 3.2 $\times$ 10$^{11}$ M$_{\odot} $ with scale length a$_{*}$ = 3.8 kpc provides the best compromise. Figure \ref{fig:fcmass} compares the cumulative mass profiles for our model to the observed stellar mass. Table \ref{table:fparameters} summarises the parameters used in the double Hernquist model for Fornax. Figure \ref{fig:fornaxprofiles} shows that our choices for the Fornax model lead to a good overall agreement with the ICM profiles to the observations at the end of the V0 and V2 simulations. For our simulated profiles, gas density is lost in the very center which is partly due to gas stripping and resolution. This gas loss would be different if cooling and heating were accurately modelled in our simulation. This is also the same for NGC 1404. Significantly however, our results do not rely on the central gas cores.

\begin{table}[h!]
\centering
\caption{\label{table:fparameters} Fornax Cluster Model Parameters}
\begin{tabular}{lc}
\hline
\hline
Double Hernquist \\
\hline
M$_{ \text{dm}}$ / 10$^{13}$M$_{\odot} $          &  6.0            \\
a$_{ \text{dm}}$ / kpc          & 250.0         \\
M$_{*}$ / 10$^{11}$M$_{\odot} $          &   3.2          \\
a$_{*}$ / kpc            & 3.8      \\
\hline
Double $\beta$ Model \\
\hline
A$_{1} $ / cm$^{-3}$        & 0.151         \\
r$_{1}$ / kpc  & 0.623           \\
$\beta_{1}$          & 0.44           \\
A$_{2} $ / cm$^{-3}$           & 0.0024         \\
r$_{2} $ / kpc & 55.0           \\
$\beta_{2}$          & 0.48           \\
\hline
\end{tabular}
\end{table}

\subsubsection{NGC 1404}
\begin{figure}[h!]
\centering
\includegraphics[scale=0.3]{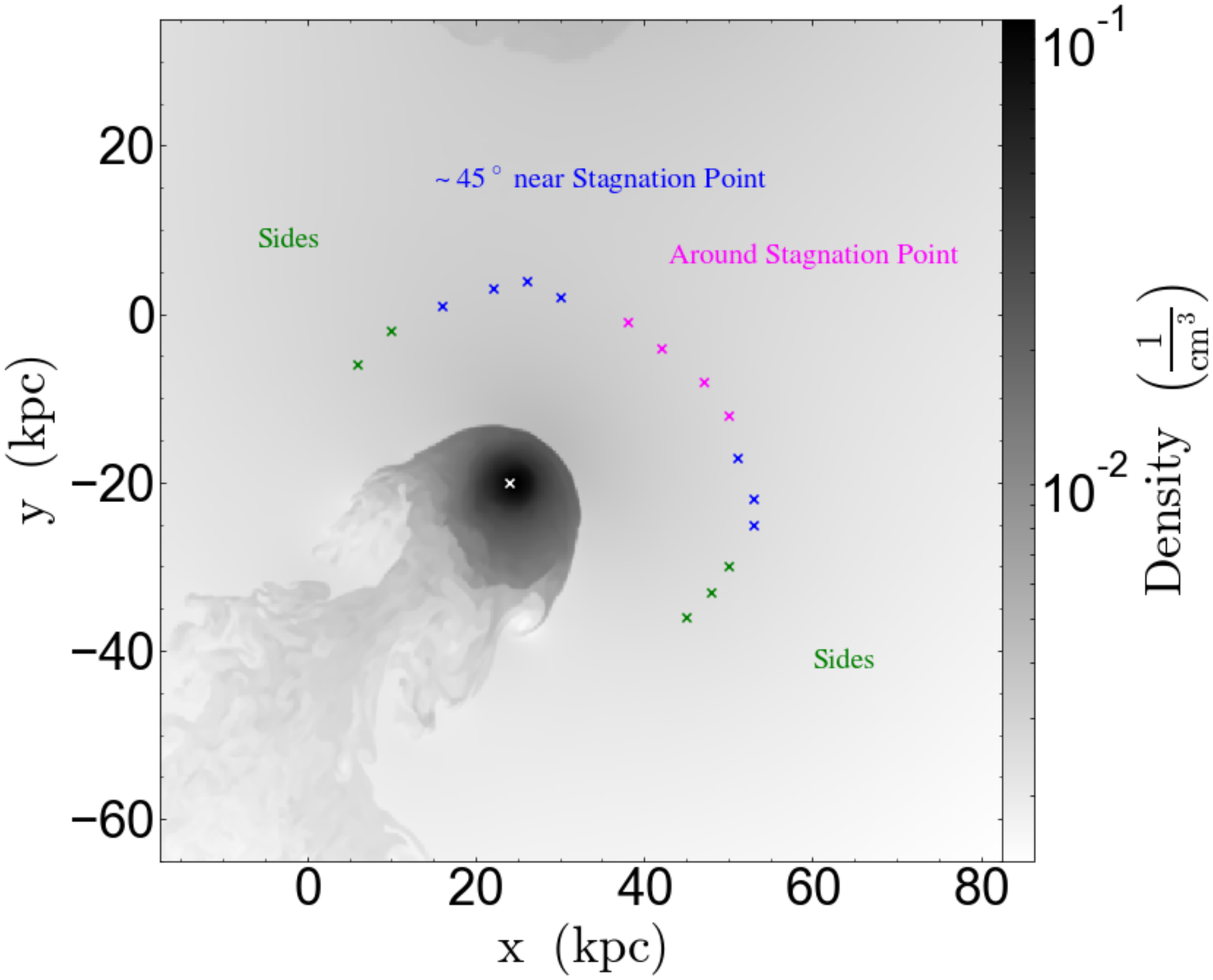}
\includegraphics[scale=0.22]{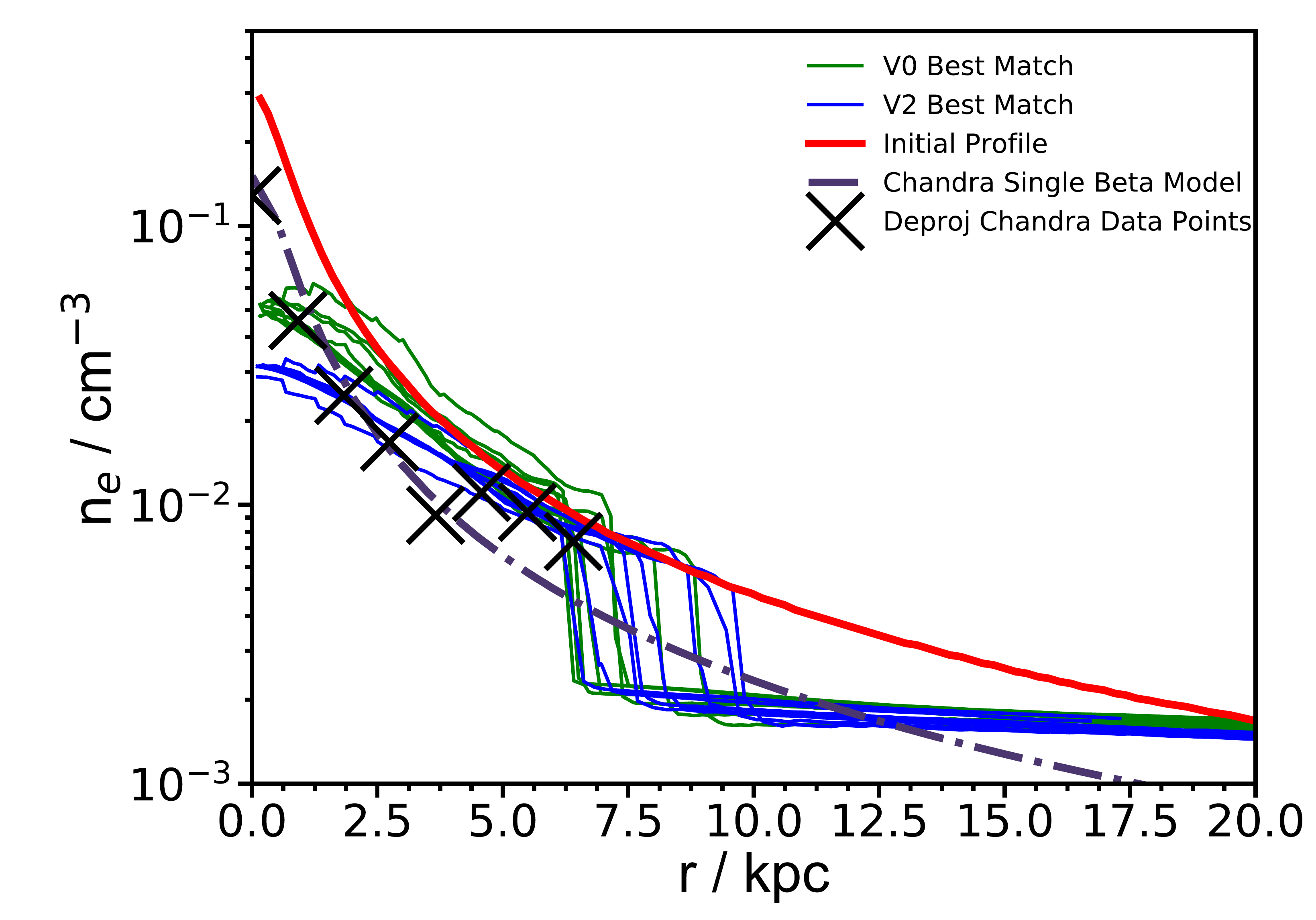}
\includegraphics[scale=0.22]{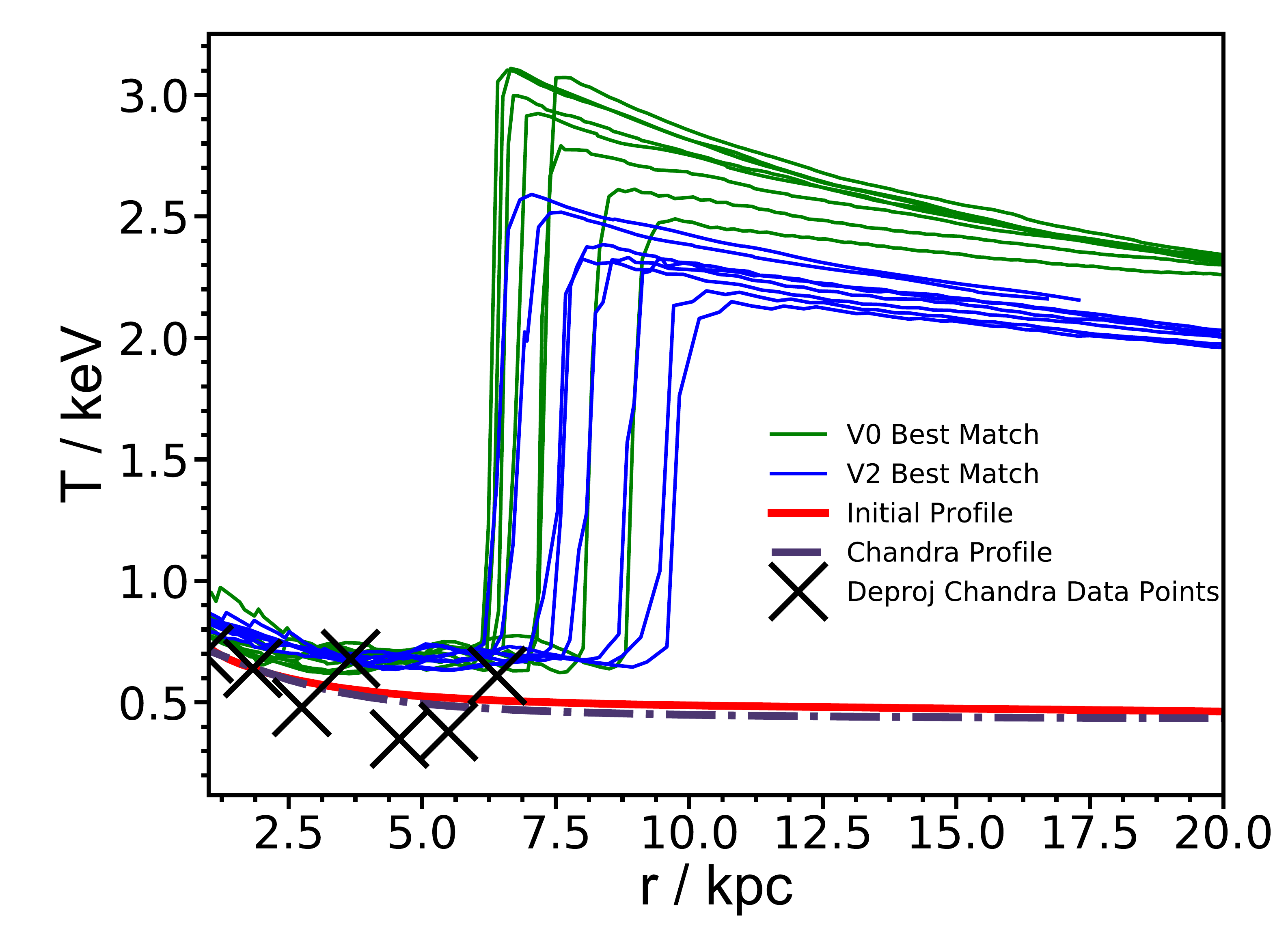}
 \caption{Comparison of simulated and observed gas atmosphere profiles for NGC 1404 from the V0 and V2 simulations. The observed profiles for electron density (middle) and temperature (bottom) are azimuthally averaged. For the simulated galaxy, profiles are taken along rays from the galaxy center to the locations indicated in the top panel (the example shown is for the V0 simulation). The simulated profiles are taken at the stage of the merger which provide a visual match to observation. The sharp drop in density marks the radius of the galaxy, where the density then becomes the Fornax ICM. The simulations reproduce the observed upstream stripping radius and agree well with the observed profiles around this point considering our simple model. The decrease in density at the very centre of NGC 1404 in our profiles is partly due to gas stripping and internal sloshing, and partly due to resolution (the decrease in density gets a little lower after each pericenter passage - this is why we set the initial profiles twice as high as the observational profile).  The temperature profiles reveal an increased ICM temperature around the stripped NGC 1404 atmosphere. This is due to the stolen atmosphere effect as explained further in Section \ref{sec:sa}.}
\label{fig:n1404bestmatch}
\end{figure}

As with Fornax, we tailor the simulation to NGC 1404 using Chandra data to model its gas density as a single $\beta$ model. Figure \ref{fig:n1404bestmatch} compares the NGC 1404 gas density and temperature profiles taken from Chandra with our initial and evolved model profiles for the V0 and V2 simulations. We initially set the gas density twice as large as the Chandra single $\beta$ model as from testing we find that during the early Fornax core passages, the NGC 1404 gas density decreases by about a factor of two due to gas stripping and we need to match to observations at the evolved stage. In the course of the merger, the central gas density of NGC 1404 decreases somewhat. This is partly a real effect due to gas loss via gas stripping combined with gas redistribution via sloshing in NGC 1404, but the inner kpc is affected by resolution as well. We aimed to keep a good match to the observed gas profile outside \sim 4 kpc. Furthermore, our model choice of potentials, gas content, and galaxy orbit which are driven by the overall properties of Fornax and NGC 1404, lead to an upstream stripping radius of \sim 6 kpc, remarkably close to the observed 8 kpc, given our simple model. \par

\begin{figure}[h!]
\centering
\includegraphics[scale=0.42]{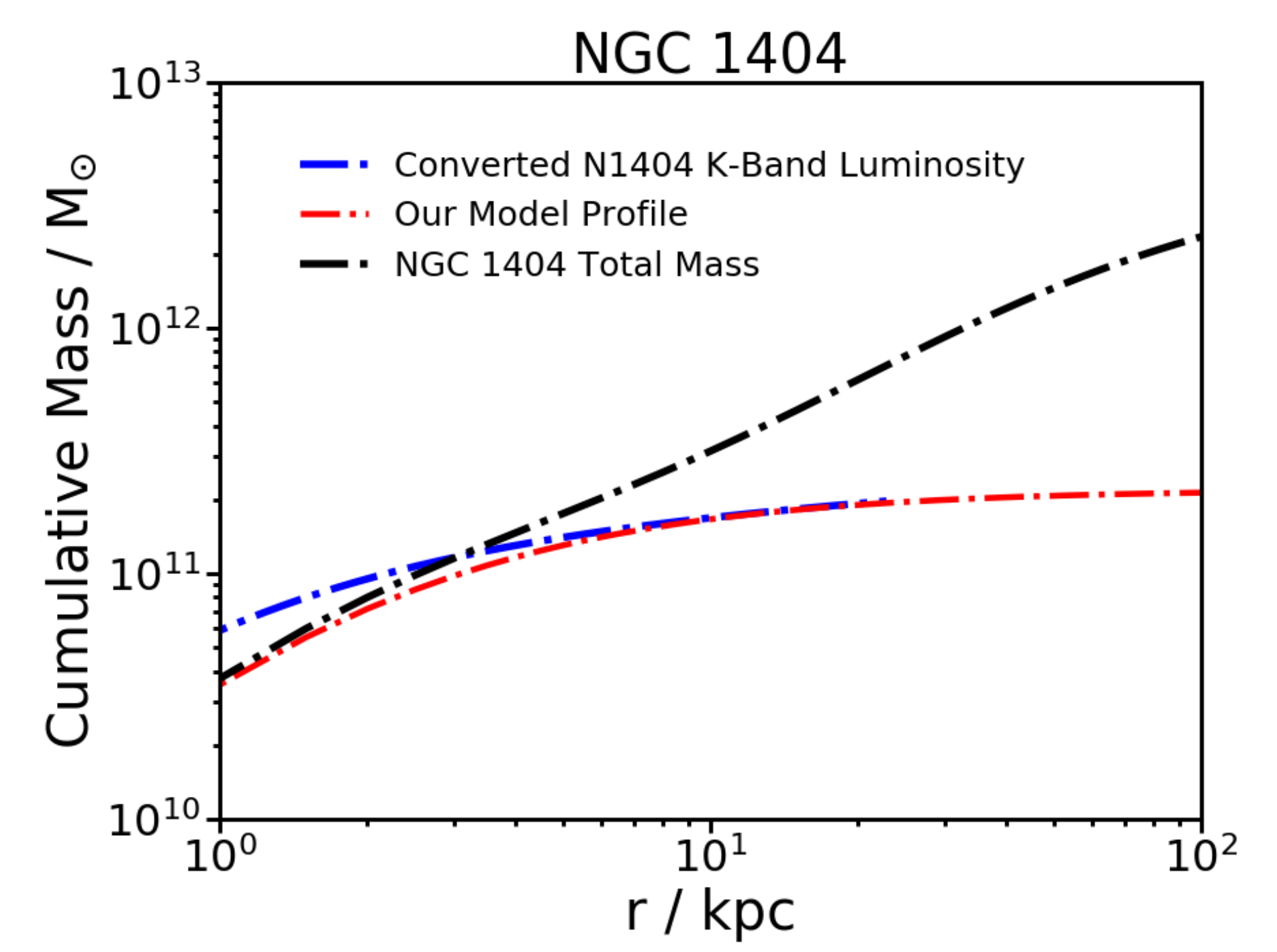}
\caption{Cumulative mass profile for our model NGC 1404 compared to cumulative stellar mass of NGC 1404, converted from K-Band luminosity. The total mass of the model galaxy consists of an inner and outer Hernquist potential, where roughly the inner potential represents the NGC 1404 stellar component. As a compromise between the stellar light data and the observed NGC 1404 temperature profile, We chose an inner component somewhat less compact than observed because otherwise the hydrostatic initial setup of the cluster results in an unrealistically high central NGC 1404 temperature.}
\label{fig:nkband}
\end{figure}

Again, like for Fornax, we use a double Hernquist potential to model the overall potential of NGC 1404. To constrain the inner components of NGC 1404 we are guided by the K-band luminosity data (see Figure \ref{fig:nkband}) and fit M$_{*}$ and a$_{*}$ appropriately with values of M$_{*}$ = 2.2 $\times$ 10$^{11}$ M$_{\odot}$ and a$_{*}$ = 1.5 kpc respectively. Here we also find the compromise between matching the stellar data and keeping a reasonable gas temperature profile. For the outer potential component, we follow the model of M89 used in \citet{Roediger2015c} as M89 is comparable to NGC 1404 in regards to the gas temperature and stellar luminosity. Therefore we select values of M$_{\text{dm}}$ = 0.45 $\times$ 10$^{13}$ M$_{\odot}$ and a$_{\text{dm}}$ = 45.0  kpc. Table \ref{table:nparameters} summarises the parameters for the double Hernquist model and the single beta model parameters for NGC 1404. 

\begin{table}[h!]
\centering
\caption{\label{table:nparameters} NGC 1404 Model Parameters}
\begin{tabular}{lc}
\hline
\hline
Double Hernquist     \\
\hline
M$_{ \text{dm}}$ / 10$^{13}$M$_{\odot} $          &    0.45      \\
a$_{ \text{dm}}$ / kpc            & 45.0          \\
M$_{*}$ / 10$^{11}$M$_{\odot} $          & 2.2            \\
a$_{*}$ / kpc           & 1.5      \\
\hline
Single $\beta$ Model \\
\hline
A$_{1} $ / cm$^{-3}$          & 0.151         \\
r$_{1} $ / kpc & 0.623           \\
$\beta$          & 0.5           \\
\hline
\end{tabular}
\end{table}

\subsection{Metallicity Profiles}
In our simulation, we include a mass scalar which holds metallicity information for both NGC 1404 and Fornax. This mass scalar is carried throughout the simulation as a dye and does not interact with the problem dynamics. In turn, this allows us to track the redistribution of metals throughout the cluster during the merger. For Fornax, the Chandra and XMM data are valid out to 250 kpc; from this point we extrapolate the observed data to approach a metallicity of 0.3 solar as observed in other clusters \citep{Simionescu2011a}. Likewise for NGC 1404, the data are only valid for the inner 8 kpc, so we adapt again so that it approaches 0.3 solar at large radii. The abundance profiles we use are:

 \begin{equation} \mathrm{Fe_{\text{Fornax}}} = 0.42 \bigg( 1+ \bigg( \frac{\text{r}}{22\:\text{kpc}}\bigg) \bigg)^{-0.5} +0.28 \end{equation} 

 \begin{equation} \mathrm{Fe_{\text{N1404}}} = 0.45 \bigg( 1+ \bigg( \frac{\text{r}}{12\:\text{kpc}}\bigg) \bigg)^{-0.25} +0.2 \end{equation} 

\begin{figure}[h!]
\centering
\includegraphics[scale=0.36]{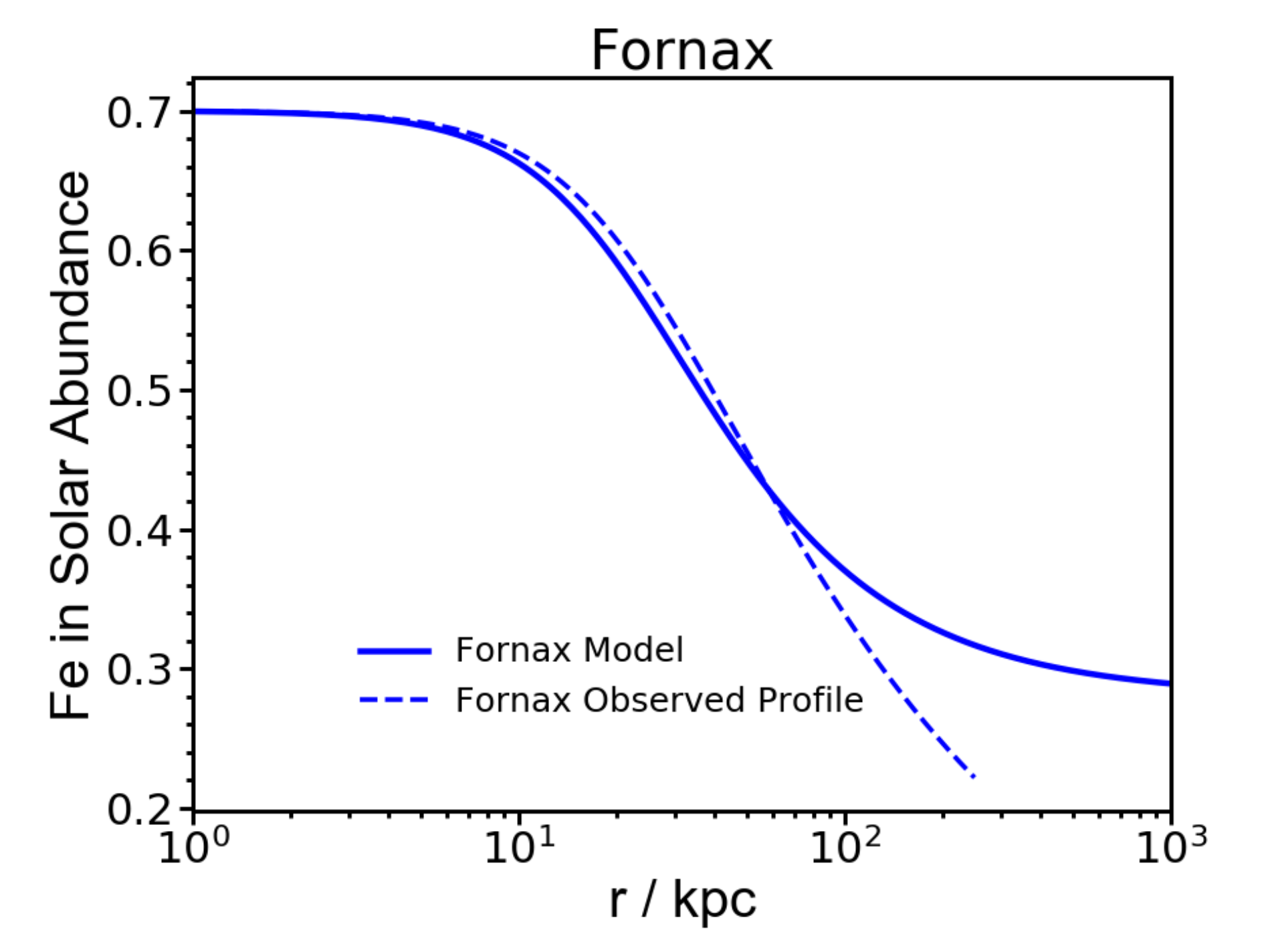}
\includegraphics[scale=0.36]{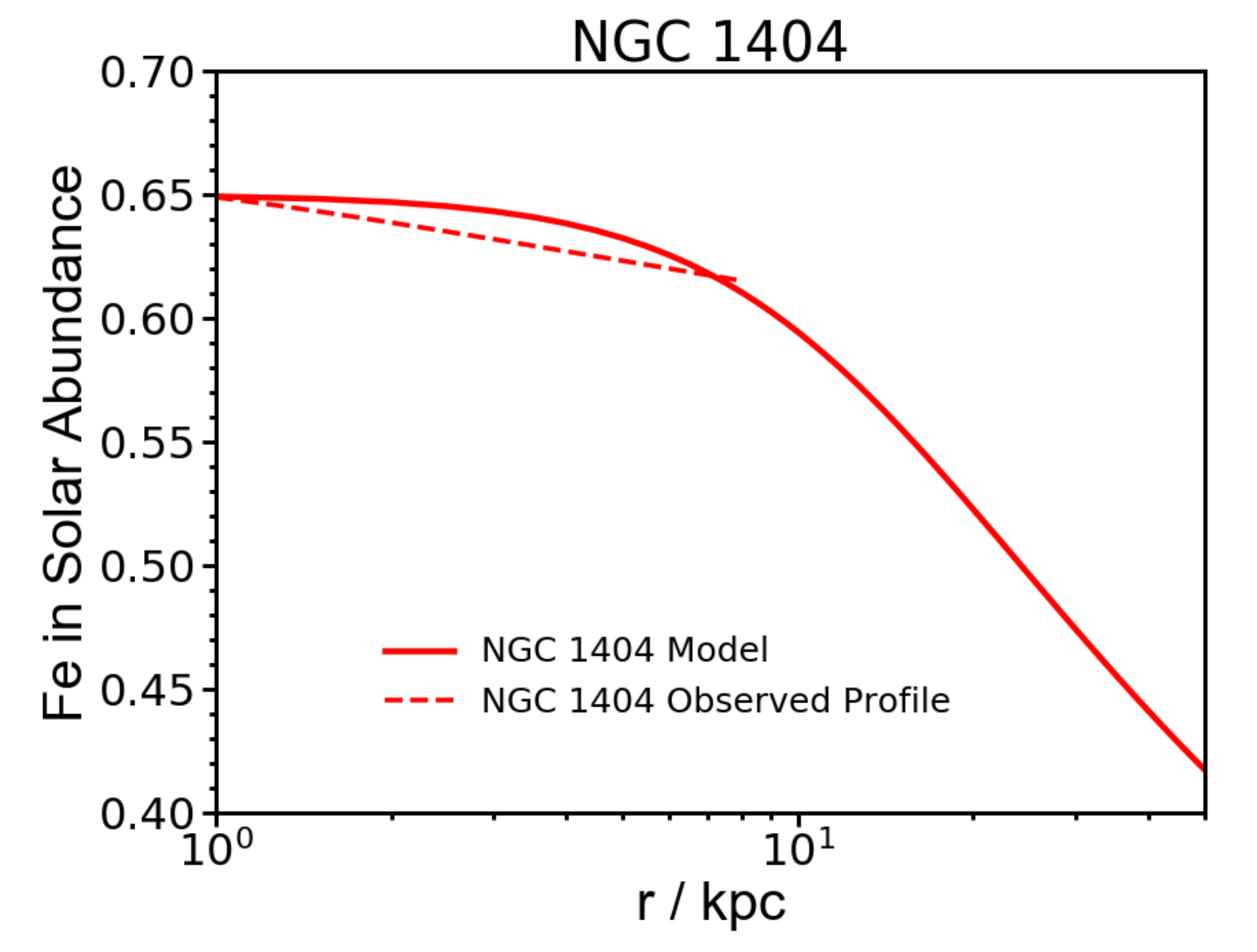}
\caption{Top: Initial metallicity profiles in Fe solar abundance for Fornax. Chandra and XMM data are valid out to 250 kpc for Fornax, thus we extrapolate the observed data to approach a metallicity of 0.3 solar as observed in other clusters \citep{Simionescu2011a}. Bottom: Initial metallicity profiles in Fe solar abundance for NGC 1404. Chandra data is only valid for the inner 8 kpc of the galaxy, so again we extrapolate so that it approaches 0.3 solar at large radii. Observed profiles are represented by the dashed lines and our model profile by the thick line.}
\label{fig:metpro}
\end{figure}

Figure \ref{fig:metpro} compares the model and observed profiles.

\subsection{FLASH - The Simulation Code}
\label{sec:simcode}
Our simulations use the FLASH Code - a high performance modular code developed to handle multi-physical problems (\citealp{Fryxell2000}). We utilise its 3D hydrodynamic + Nbody capabilities to simulate the interaction between the gaseous and collisionless (dark matter and stars - both self gravitating) components, respectively. Including the self gravity of the collisionless particles, and the gas, as well as the gravity between both components, allows us to accurately characterize tidal forces and dynamical friction during the merger. This has a significant impact on the orbit of the galaxy and hence plays a pivotal role in the merger timescales. \par

For simulation V0, we use a particle resolution of 4 million particles for Fornax and 400,000 particles for NGC 1404. Whereas for V1 and V2, we use a particle resolution of 1 million particles for Fornax and 100,000 particles for NGC 1404. FLASH Code has the capability to use adaptive mesh refinement defined by specific refinement criteria. Our simulation refines a grid block if it has more than 200 particles. This simple choice allows us to capture the tail of the galaxy as well as the centers of both the galaxy and the cluster at a maximum resolution, since these regions contain the largest particle numbers. At the highest refinement level the resolution is 0.22 kpc. The Fornax center and the NGC 1404 are refined to this level typically out to 50 kpc and 12 kpc respectively. This refinement criteria certifies that the majority of the wake of NGC 1404 is also refined to this highest level. We found that not refining enough i.e. having a lower resolution, produced ripples in the potential centers which expanded outwards from the centers of the cluster and galaxy. This also caused the very centers of galaxy and cluster to lose gas. Both effects decrease with higher resolution and are negligible in our simulations.

\subsection{Simulation Design}
We select a simulation box size of 1.5 Mpc$^{3}$. This is chosen to be sufficient to capture the potential of the Fornax gas past the virial radius, \sim 750 kpc, and so the infall path is not subjected to boundary effects. The simulation boundaries are set to outflow (shocks can leave the simulation domain) but we do not anticipate any impact on the simulation as the main physics occurs far enough away from the boundaries. Our simulations neglect both radiative cooling and heating by active galactic nuclei (AGN), i.e., we assume that over time, both processes balance out as implied by observations (\citealp{Birzan2008}). The addition of cooling, and the balancing AGN activity, would affect mostly the gas properties in the central kpc's of NGC 1399, and the very central kpc's of NGC 1404, where the cooling time is shortest. None of our results rely on the central gas cores.

\subsubsection{Orbit}
\label{sec:orb}
We match the velocity of NGC 1404 from observational constraints by \citet{Machacek2005a}, \citet{Scharf2005a} and \citet{Su2017f}. This must be reflected in our simulation. Further, we match the radius of our model galaxy to the radius from the center of NGC 1404 to the upstream cold front measured in \citet{Machacek2005a} and \citet{Su2017f} of $\sim$ 8 kpc. \par

At the beginning of the simulation, Fornax sits at the center of the simulation box with NGC 1404 placed at roughly the virial radius of the cluster, \sim 750 kpc. We test several orbits for NGC 1404 through Fornax (Table \ref{tab:sim_runs}). In simulation V0, we give NGC 1404 only an initial tangential velocity to control the impact parameter of the merger.  A head on merger would strip all gas from NGC 1404 which is ruled out by observations. An initial tangential velocity of 150 km s$^{-1} $ in the -y-direction leads to a gas stripping radius comparable to the observed one. However, such an almost zero-velocity infall from the virial radius is not what is typically seen in cosmological simulations. \citet{Vitvitska2002} show that during the initial phase of a merger, the merging subcluster already has an infall velocity of $\sim$ 1.1v$_{c}$ when it reaches the virial radius, where v$_{c}$ is the circular velocity of the main cluster defined in the Hernquist model by v$_{c}$ = $\mathrm{\frac{\sqrt{GM(r)r}}{r + a}}$, where a is the scale length of the main cluster. Their work also finds that, for a major merger, the tangential component of the infall velocity is typically around 0.45v$_{c}$ whereas for a minor merger this typical value becomes 0.71v$_{c}$. We therefore run two further simulations of a faster infall, V1 and V2 described in Table \ref{tab:sim_runs}. Using a virial radius of 750 kpc, a scale length of 250 kpc and a virial mass of 3.4 $\times$ 10$^{13}$ M$_{\odot}$ (calculated using the Hernquist mass formula given in eq.1) gives a circular velocity of v$_{c}$ = 332 km s$^{-1}$ and therefore an infall velocity of v$_{in}$ = 365 km s$^{-1}$. Simulations V1 and V2 have the same v$_{in}$, but different tangential and radial velocities, v$_{\perp}$ and v$_{\parallel}$ respectively (see Table \ref{tab:sim_runs}), sampling the range given by \citet{Vitvitska2002}.

\begin{deluxetable}{ccccc}
\tablecaption{Simulation Runs\label{tab:sim_runs}}
\tablewidth{0pt}
\tablehead{
\colhead{Sim Name} & \colhead{Initial Separation} & \colhead{v$_{\parallel}$} & \colhead{v$_{\perp}$} \\
\colhead{} & \colhead{(kpc)} & \colhead{(km s$^{-1}$)} & \colhead{(km s$^{-1}$)}}
\startdata
V0   & 650 (x = -430, y = 480)       & 0                    & +150                   \\
V1   & 750 (x = 750, y = 0)            & -333               &  0.45v$_{c}$ = -149   \\            
V2   & 750  (x = 750, y = 0)           & -279               &  0.71v$_{c}$ = -236  \\                  
\enddata
\end{deluxetable}

It turns out that in simulations V1 and V2 the galaxy reaches about the virial radius on its first apocenter passage, i.e. it falls back into the Fornax cluster almost from rest after that point. This means that our simulation V0 is comparable to the second and third infall in V1 and V2. We therefore orient our simulations such that their final infall occurs in a comparable direction; i.e. in simulation V0, NGC 1404 approaches from the top left (NE). In V1 and V2, NGC 1404 approaches from the +x-direction, from the right (W), with a tangential velocity component in the -y-direction.

\section{Results}
\label{sec:results}

\subsection{Overall Merger History}
\begin{figure*}
\centering
\includegraphics[scale=0.75]{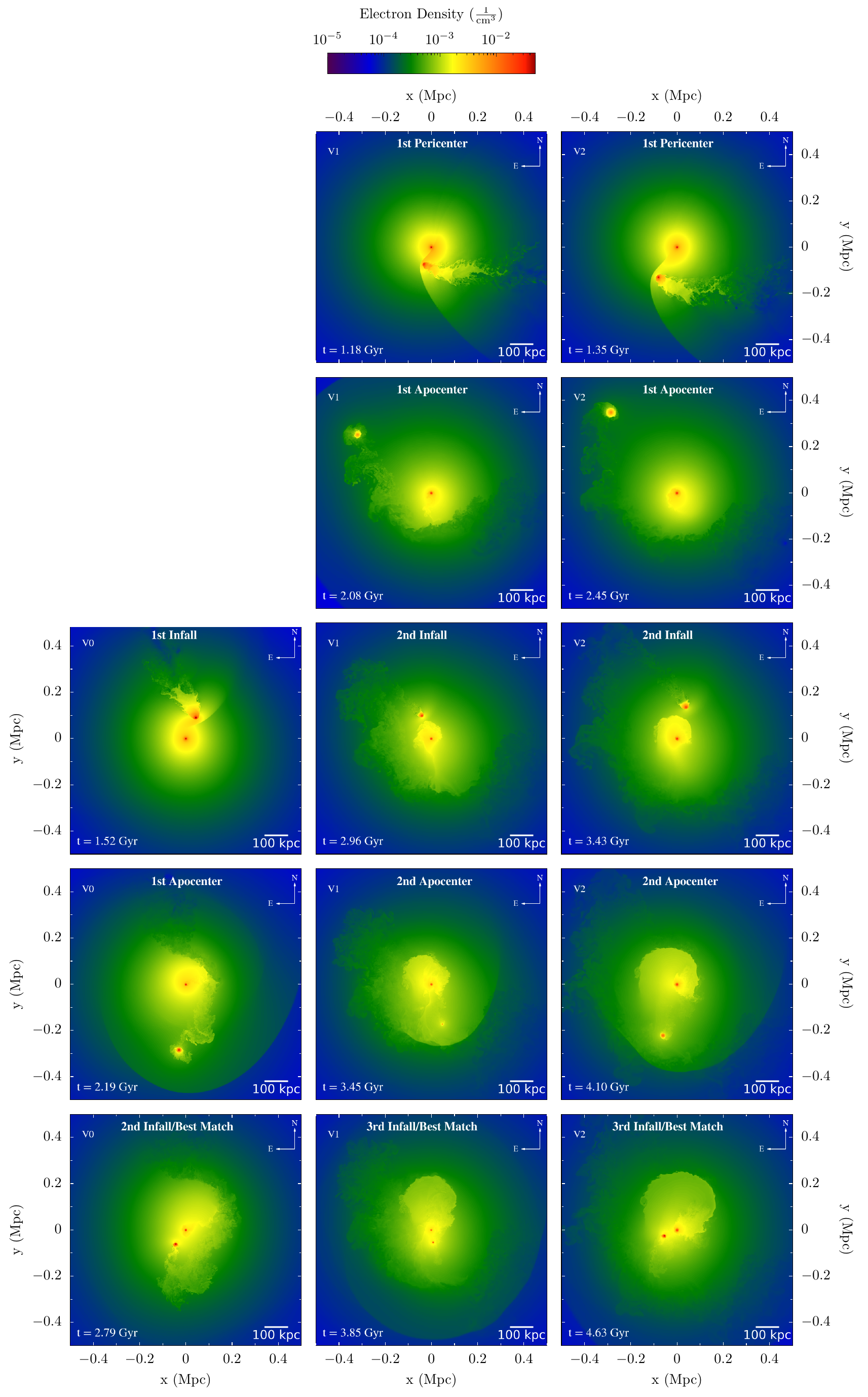}

\caption{Electron density snapshots in the orbital plane for the overall evolution of the infall of NGC 1404 into Fornax. Each column represents a different simulation run. From left to right; V0, V1, V2. In rows, we align the snapshots by evolutionary stage rather than exact time. Row 1 shows the first infall of NGC 1404 into Fornax for V1 and V2 revealing the bow shock and a long gas tail of NGC 1404. Row 2 shows the galaxy now at apocenter for V1 and V2 revealing irregular flow patterns. Row 3 shows the second infall of V1 and V2 along with the first infall of V0. Here, for V1 and V2, the galaxy hosts a short gas tail but is still moving supersonically evidenced by the bow shock. For V0, this first infall is comparable to the second infall of V1 and V2 as the galaxy begins its infall with almost zero velocity. Row 4 shows the 1st apocenter of V0 and 2nd apocenter of V1 and V2. The detached bow shock at this stage is clearly evident, propagating outwards south of the galaxy. Row 5 represents the best match for each simulation. At this stage the cold front to the north of the cluster center, along with the sharp upstream edge, truncated atmosphere, stripping radius and faint tail of NGC 1404 are comparable to the observation. For the electronic version of the paper, we include an electron density animation of the three simulations alongside each other in separate panels in the same fashion as this figure. The animation covers the full run time for each simulation. At the start, simulations V1 and V2 begin to run. V0 begins to run once V1 and V2 reach their first apocenter. This is because the first infall of V0, which has an almost zero initial velocity, is comparable to the second infall of V1 and V2.}
\label{fig:wholeevo}
\end{figure*}

\begin{figure*}
\centering
\includegraphics[scale=0.84]{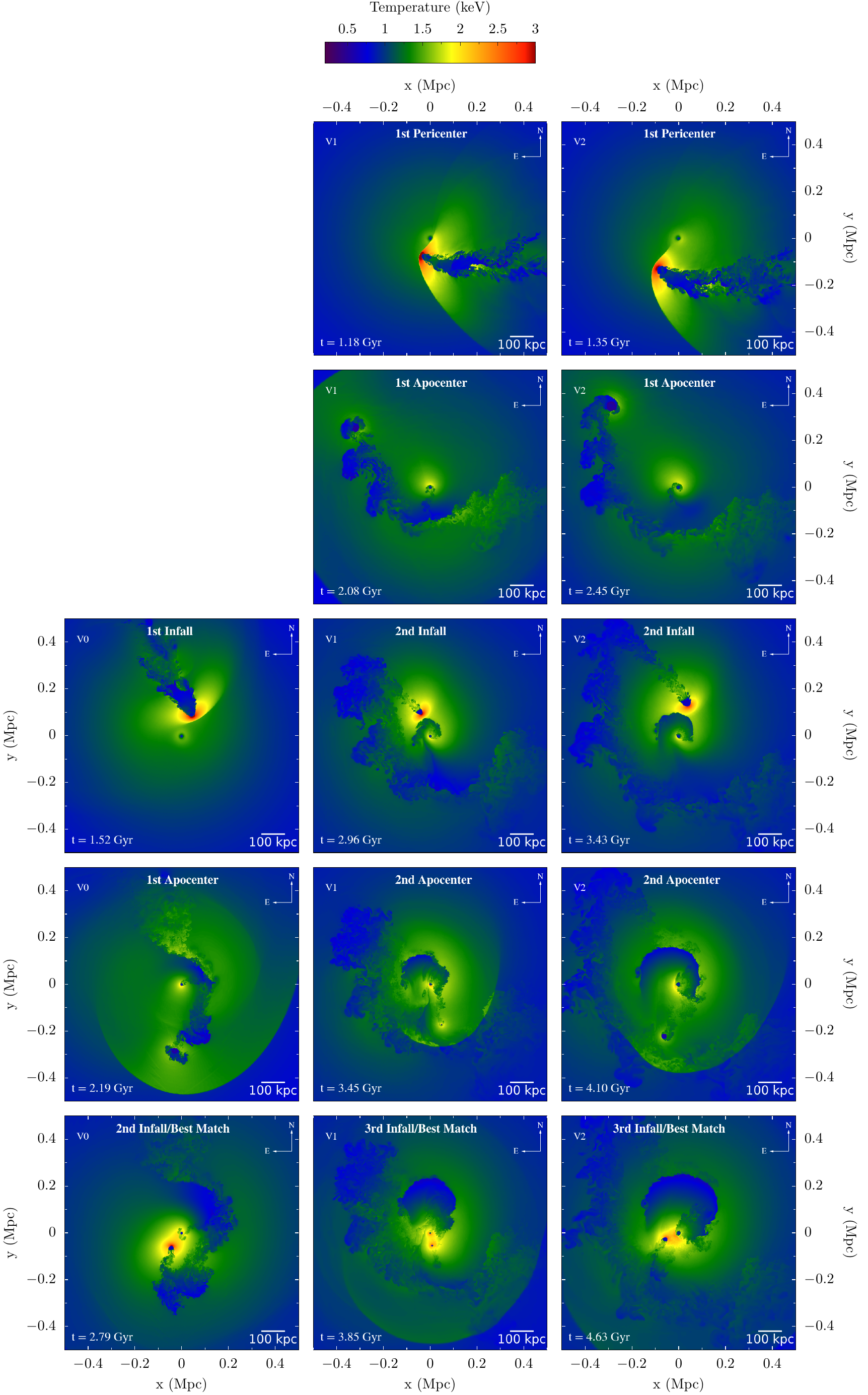}

\caption{Same as the previous figure but temperature slices in the orbital plane for the overall evolution of the infall of NGC 1404 into Fornax. Sloshing fronts and NGC 1404's turbulent wake are seen more easily in the temperature slices.}
\label{fig:wholetempevo}
\end{figure*}

The overall merger history for all three simulations is presented in Figures \ref{fig:wholeevo} and \ref{fig:wholetempevo} using snapshots of electron density and temperature slices, respectively, in the orbital plane. We first describe simulation V0. Here, NGC 1404 approaches from the NE and reaches the center point on its first infall into Fornax at an epoch of 1.52 Gyr. At this stage, the galaxy has a long extended gas tail \sim 100 kpc in length, moving supersonically, as evidenced by the strong bow shock in front of the galaxy. Moving past pericenter, the galaxy is slowed by gravity and dynamical friction as it approaches apocenter producing a slingshot like tail. Apocenter is attained at 2.19 Gyr, revealing highly irregular flow patterns around the galaxy leaving behind a wake of turbulent gas. Moving past apocenter, the galaxy begins to infall into Fornax a second time reaching a match to observation at an epoch of 2.79 Gyr, 1.27 Gyr after its last pericenter passage. At this second infall stage, the sharp upstream edge, truncated atmosphere and stripping radius of the galaxy are comparable to the observation. The passage of NGC 1404 near the Fornax center triggers sloshing in the central Fornax ICM which leads to a cold front north of the Fornax center as observed.

The simulations V1 and V2 portray an almost identical picture for the recent merger history, but they include one earlier Fornax core passage. NGC 1404 originally approaches from the W and reaches its first pericenter south of Fornax at 1.18 Gyr and 1.35 Gyr for V1 and V2 respectively before moving out to the apocenter in the NE. Again, NGC 1404 still has a prominent gas tail on first infall, despite the higher infall velocity and it is not until approaching the first apocenter that NGC 1404 is fully truncated. \par

This first encounter induces typical sloshing motion in Fornax, producing a prominent front which sweeps around the center (Rows 2 and 3 in Figures \ref{fig:wholeevo} and \ref{fig:wholetempevo}). The prominent sloshing front in the V2 simulation reveals the appearance of KHI rolls along its interface. The KHIs are clearest in V2 because here the shear flow along the cold front is cleanest and strongest. The second infall occurs from the NE along the NW of the Fornax core at 2.96 Gyr for V1 and 3.43 Gyr for V2. This stage corresponds to the first infall in V0 except for the gas contents of NGC 1404. Critically during the second infall, even though the galaxy resembles a sharp upstream edge and truncated atmosphere, it is still moving supersonically, as evidenced by a bow shock, and it is not until the third infall that this bow shock disappears. \par

The third infall occurs from the S/SE. At this stage, the galaxy still harbours a sharp upstream edge and truncated atmosphere. Most significantly at this stage, if we orient the images to match the observed infall direction of NGC 1404 from the SSE, the direction and position of the sloshing fronts appear to coincide with observational images of Fornax. When comparing the two test simulations with V0, the story looks relatively similar when comparing the evolution past the second apocenter. It could be considered that actually the V0 simulation also represents a third infall in the sense that this simulation begins at the second apocenter of the test simulations. This would concur as V0 starts with an almost zero infall velocity which would correspond to the galaxy reaching apocenter and falling back in again.

\subsection{Second + Third Infall}
\begin{figure*}
\centering
\includegraphics[scale=0.429]{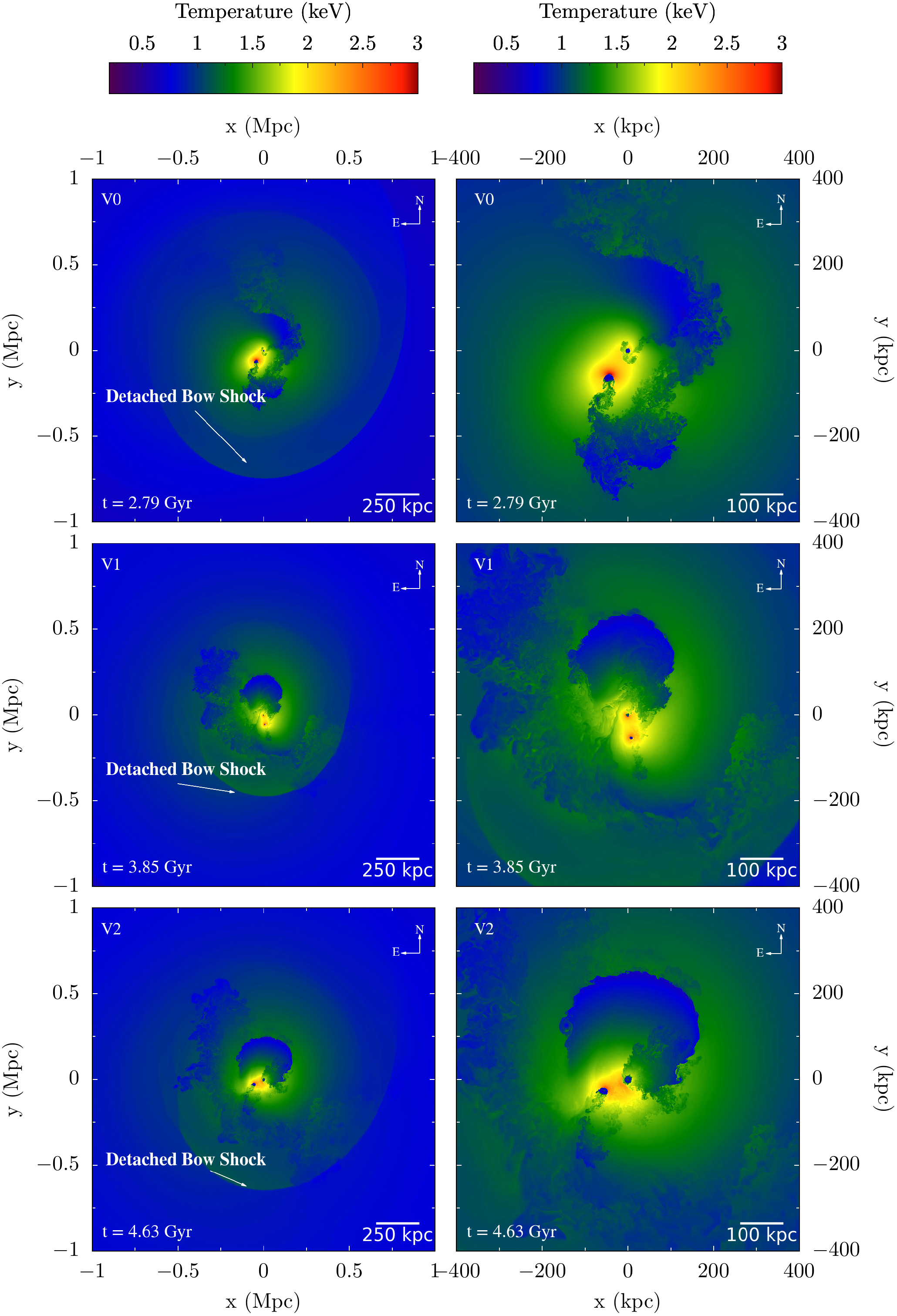}
\includegraphics[scale=0.429]{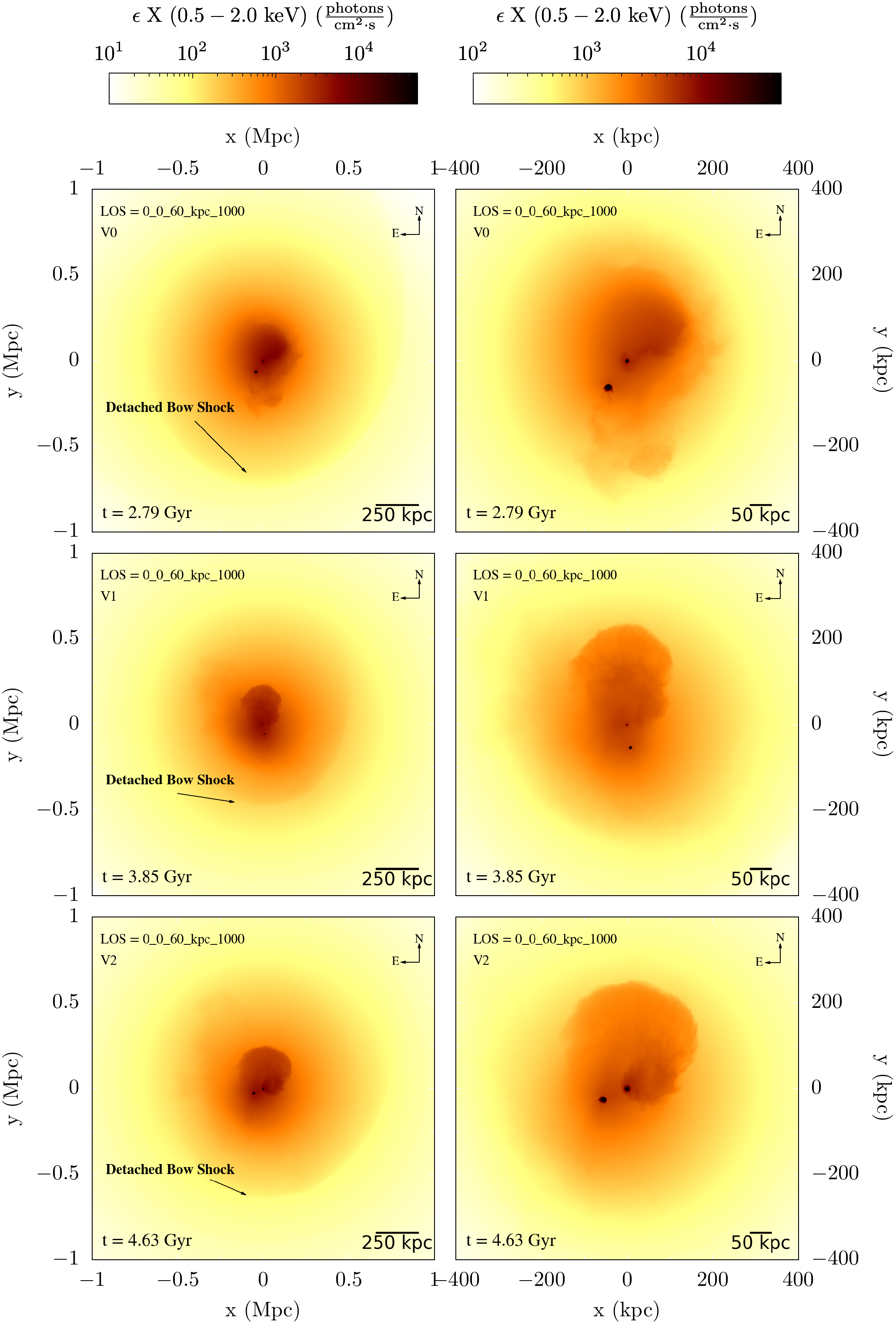}
\caption{Comparison of the best match stage for all 3 simulations, in temperature slices in the orbital plane (columns 1 and 2) and mock X-ray images (columns 3 and 4). Columns 1 and 3 show the cluster-wide view, whereas columns 2 and 4 show a zoomed in image of the cluster center. For the cluster wide images, we indicate the position of the detached bow shock south of NGC 1404 which is a remnant of the previous infall of NGC 1404. In the temperature slices, the path of destruction caused by the merger of NGC 1404 is revealed in the cluster gas, and we can see where the galaxy did not at least stir turbulence (other galaxy's may have). In the zoomed in temperature slices, the stolen atmosphere effect is evident surrounding the halo of the galaxy where the temperature is increased higher than the original ICM temperature. The X-ray photon emissivity projections are calculated in the 0.5 - 2.0 keV energy band along the axis perpendicular to the orbital plane. The X-ray images show a prominent cold front to the north of the cluster center matching the XMM image of the cluster (see Figure \ref{fig:obsxmm}) along with the truncated atmosphere and faint gas tail of NGC 1404.}
\label{fig:temp_xray}
\end{figure*}

As shown in Figures \ref{fig:wholeevo} and \ref{fig:temp_xray}, in regards to the V0 simulation, the second infall at an epoch of 2.79 Gyr provides the best match to observation for this simulation. Here as already mentioned, the simulation reproduces the sharp upstream edge and truncated atmosphere with a galaxy radius between 7 - 11 kpc compared to the observationally measured radius of 8 kpc (\citealp{Machacek2005a}, \citealp{Su2017f}). By using our simple method of prescribing appropriate gravitational potentials, gas contents and using a sensible orbit through Fornax, we are able to attain the correct radius (see also Figure \ref{fig:n1404bestmatch}). At this stage, the velocity of NGC 1404 relative to NGC 1399 (the Fornax center) is 871 km s$^{-1}$.

After the merger, the temperature profile for NGC 1404 comes out a little high (see Figure \ref{fig:n1404bestmatch}) as we set up our initial model to match the observed temperature profile. At the initial stage, NGC 1404 is in the outskirts of the Fornax cluster, where the ambient ICM pressure is low. In the final timestep, it is near the Fornax center, where the higher ambient ICM pressure somewhat compresses the NGC 1404 atmosphere, leading to a slight temperature increase. In addition, the temperature profiles show that inside the atmosphere of the galaxy there are small "bumps" in temperature - these are sloshing cold fronts inside the galaxy. \par

If we would want to match the second infall of simulations V1 and V2 to the observations, we would need to invert the merger geometry about the y axis. However, despite the fact that in all cases the galaxy has a sharp upstream edge, a truncated atmosphere with a comparable radius, the galaxy is still moving too fast as seen by the bow shock. Also at this stage, the sloshing in Fornax does not quite match the observed XMM image of the cluster in terms of the prominent front north of the cluster center as seen in Figure \ref{fig:v2_xmm}. The most prominent cold front would be to the south instead. Though in each simulation the cold region does exist to the north of the cluster center, it is more like a cold fan rather than a cold front which would give the sharp discontinuity across the edge like the XMM image.\par

It is not until the third infall stage that the tail of the galaxy in the V1 and V2 simulations has almost disappeared and is now comparable to the second infall/best match of V0. A significant factor in producing this short length of tail is that our simulations are inviscid, therefore any inclusion of viscosity would likely make it difficult to replicate this feature. One point of note is that in the V0 simulation, the galaxy has a more visible short gaseous tail whereas it is much fainter in V1 and V2. This is partly due to the longer amount of time the galaxy has travelled through Fornax, ensuring the NGC 1404 outer atmosphere is fully stripped. In the rapidly varying Fornax ICM, NGC 1404 can drag part of its outer atmosphere in a tail beyond its first apocenter. \par

At the best match stage, during the third infall of the V1 and V2 simulations, NGC 1404 has a velocity relative to NGC 1399 of 603 km s$^{-1}$ and 656 km s$^{-1}$ respectively. The most noticeable difference between V1 and V2 is that NGC 1404 has lost a lot more gas in V1. This is due to the second passage of Fornax smashing right through the middle of a cold front resulting in the galaxy being heavily stripped, whereas in V2, the galaxy only clips the edge of the front due to its wider orbit. This would be evidence that the first passage through Fornax had a large impact parameter, as was suggested by \citet{Machacek2005a}.\par

The overall merger history that emerges is this:
\begin{itemize}
 \item 1.8 - 2.8 Gyr ago, NGC 1404 was at virial radius from the Fornax center to the E, NE or N, and started falling into Fornax with an almost only tangential velocity. Starting from the E would require a stronger tangential velocity than starting from the N. 
 \item This was either a particularly slow first infall, or the second infall. 
 \item1.1 - 1.3 Gyr ago, NGC 1404 passed close to the Fornax center in the NW reached its most recent apocenter 0.4 - 0.6 Gyr ago at 180-230 kpc S-SSE from the Fornax core and is now close to the next pericenter passage. At this stage the simulations reproduce a heavily stripped NGC 1404 with a gaseous tail and the overall sloshing geometry of Fornax. 
 \end{itemize} \par

As is the case in all the simulations, in terms of the best match, that being the second infall of V0 and the third infall of V1 and V2, the position of the galaxy does not come out right. To fix this would require quite some fine tuning of setting an appropriate initial velocity of the galaxy and hence its orbit through Fornax so that the galaxy would align with observed images. Crucially however, even with our small set of simulations with varied velocities and orbits, we see that the major features of NGC 1404 as well as the sloshing in Fornax are reproduced in each simulation, indicating their robustness. Therefore we believe that correcting for the position NGC 1404 would not significantly change the overall merger history.

\begin{figure}
\centering
\includegraphics[scale=0.46]{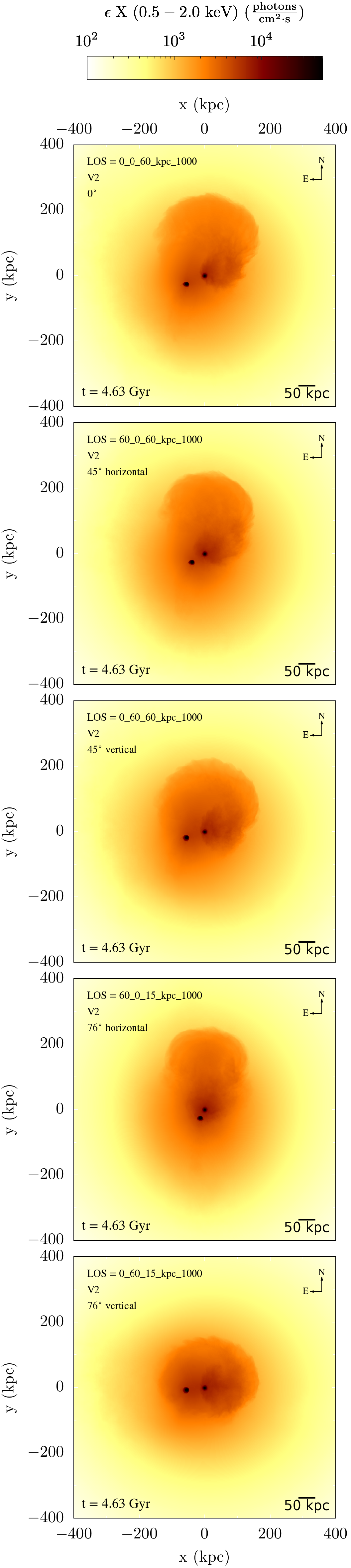}
\caption{X-ray photon emissivity projections for the V2 simulation calculated in the 0.5 - 2.0 keV energy band. The top image is a projection perpendicular to the orbital plane. The following images are a selection of LOS rotated vertically or horizontally to the orbital plane by 46$^{\circ}$ and 76$^{\circ}$. Each image is annotated with its corresponding rotation and angle. Regardless of the choice of LOS, there is no obvious difference in the brightness or shape of NGC 1404 and the center of Fornax.}
\label{fig:v2_xray}
\end{figure}

\subsection{Evolution of Fornax}
\subsubsection{V0}
Moving our focus to the evolution of Fornax through each simulation we will first examine the V0 simulation. During the first infall of NGC 1404, the center of Fornax is drawn upwards towards the galaxy due to their gravitational interaction. As the Fornax potential is rather compact, we find that using a pericenter distance that is too small results in the Fornax center either being disrupted or losing far too much gas so it is not comparable to the observed profiles. Again, this would then suggest that the first encounter between NGC 1404 and Fornax was not a close one. After the first infall, as the galaxy moves out to the apocenter of the orbit, the center of Fornax turns around and moves back towards NGC 1404. During this stage the Fornax center is itself ram pressured stripped as it moves through the ICM resulting in the gas tail trailing to the north of the cluster center, this can be clearly seen in Figures \ref{fig:wholeevo} and \ref{fig:wholetempevo} (left column) at the t = 2.19 Gyr snapshot. This stripping of the Fornax center leads to a somewhat too low central ICM density compared to the observations (Figure \ref{fig:fornaxprofiles}). Including cooling may resolve the gas core. Moving into the second infall, in V0, a large cold fan of gas becomes prominent north of the cluster center with a large cold wake behind NGC 1404 in the south, this can be seen in further detail in Figure \ref{fig:temp_xray}. The temperature profiles for Fornax in Figure \ref{fig:fornaxprofiles} reflect the cool wake of NGC 1404 and the northern cold front or fan. \par

\subsubsection{V1 + V2}
\begin{figure*}
\centering
\includegraphics[scale=0.51]{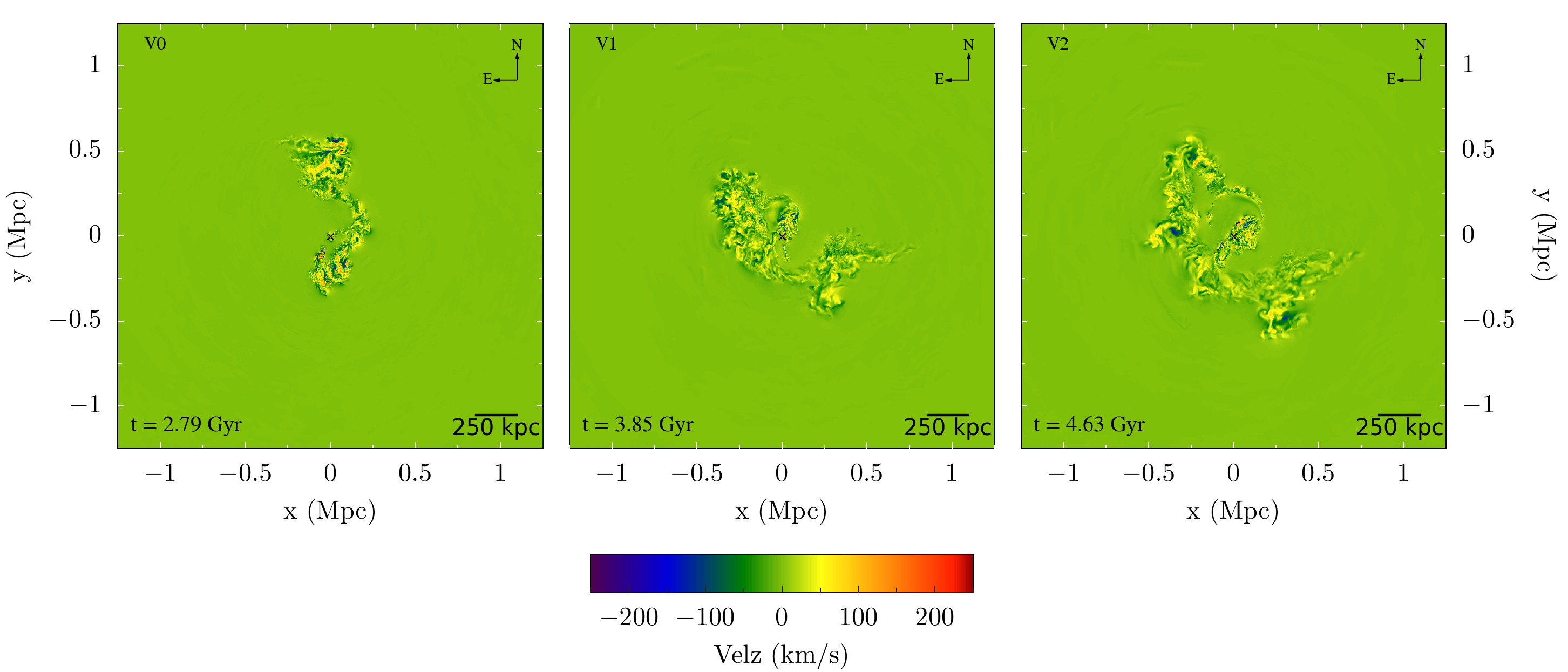}
\includegraphics[scale=0.51]{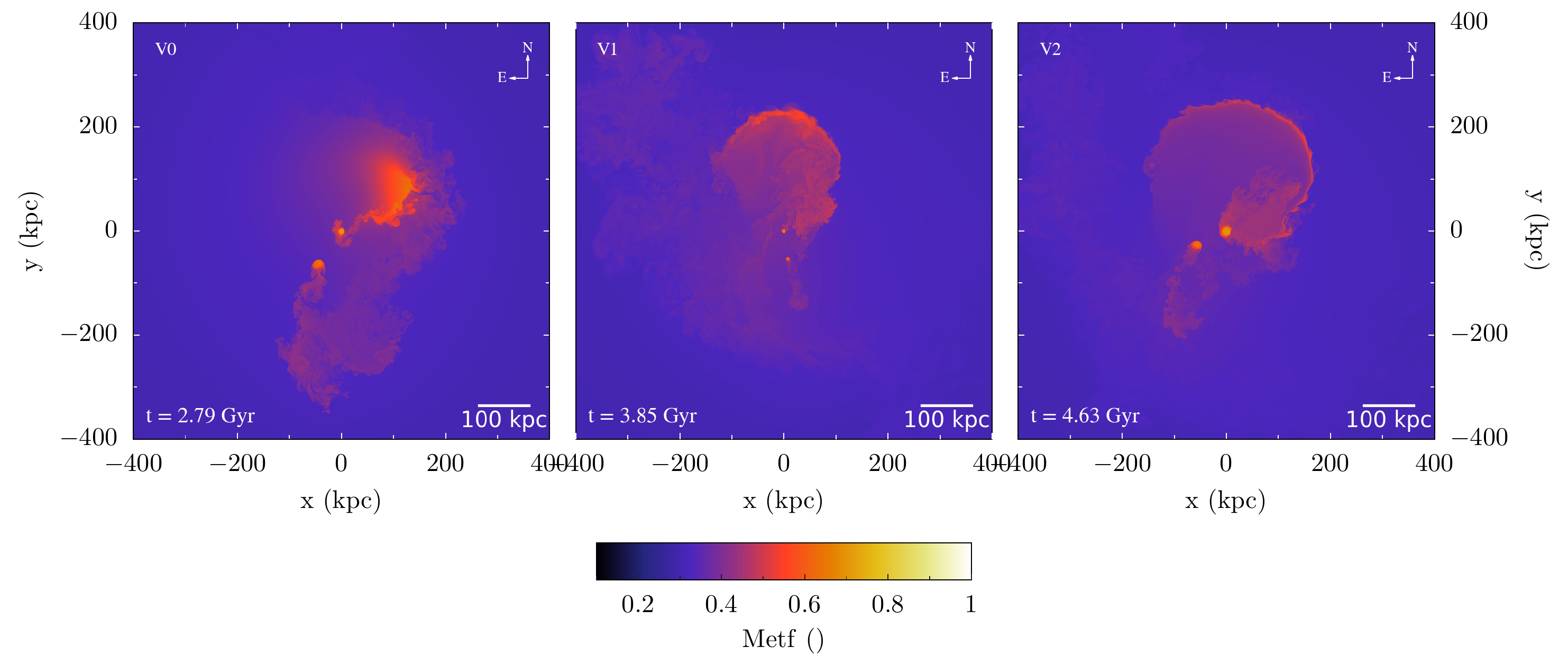}
\caption{Top: Velocity slices out of the orbital plane for each simulation revealing likely regions of turbulence caused by the infall of NGC 1404. Each column represents a simulation, from left to right: V0, V1, V2. Bottom: Same as top, but Metal fraction slices in the orbital plane in units of Fe solar abundance. The merger with NGC 1404 strips metal rich gas from the cluster center (NGC 1399) displacing it northwards producing an increased abundance in this region as was noted by \citet{Murakami2011c}.}
\label{fig:metf}
\end{figure*}

The two simulations V1 and V2 follow a similar scenario. The first encounter induces sloshing in the cluster core which has time to evolve and sweep around by the time the second encounter occurs. We can see the path of turbulence caused by the first infall of NGC 1404 across the southern region of Fornax. By the time of the third infall, the sloshing front has more time to develop, increasing in size, producing a prominent cold front north of the cluster center, along with a smaller cold front to the south as shown in Figure \ref{fig:temp_xray} (bottom panel). At the third infall stage, highlighted by the velocity map out of the merger plane shown in Figure \ref{fig:metf}, we can see the wake of destruction caused throughout the cluster due to its interactions with NGC 1404, and we can clearly see regions where NGC 1404 caused turbulence and regions where no turbulence is predicted (as a result of NGC 1404). \par

In all of the simulations, the best match case predicts a cold region of gas at temperatures around 0.9-1.0 keV to exist \sim 75-150 kpc north of the cluster center, as a result of sloshing. This cold region coincides with observations made by \citet {Murakami2011c} who detected gas of 0.9-1.0 keV in a similar region north of the cluster center. Therefore we can attribute this cold region to be the result of sloshing caused by the infall of NGC 1404. In addition, when looking at the metallicity distribution in Fornax at the second and third infall stage, shown in Figure \ref{fig:metf}, we see that there is an increased Fe abundance of \sim 0.6 Fe (solar) \sim 75-150 kpc north of the cluster center. This metal rich gas once belonged to NGC 1399 but was displaced by sloshing. \citet{Murakami2011c} detected an increased abundance of 0.53 - 0.71 Fe (solar) \sim 75  kpc north of the cluster center, which puts it roughly in the same region as our simulation. Again, we could attribute this region of increased abundance to the recent encounter with NGC 1404. \par                                         

\begin{figure*}
\centering
\includegraphics[scale=0.42]{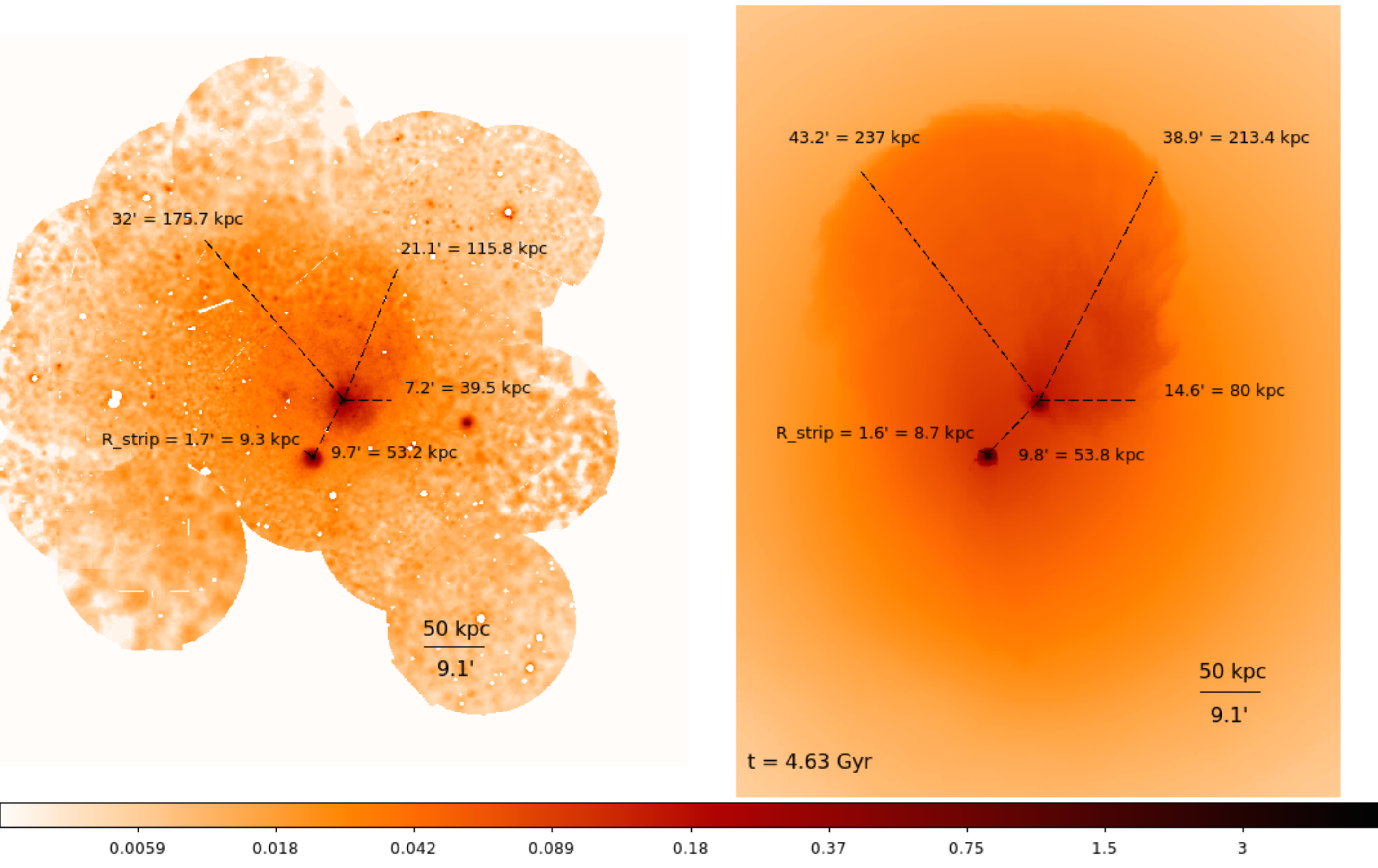}
\includegraphics[scale=0.42]{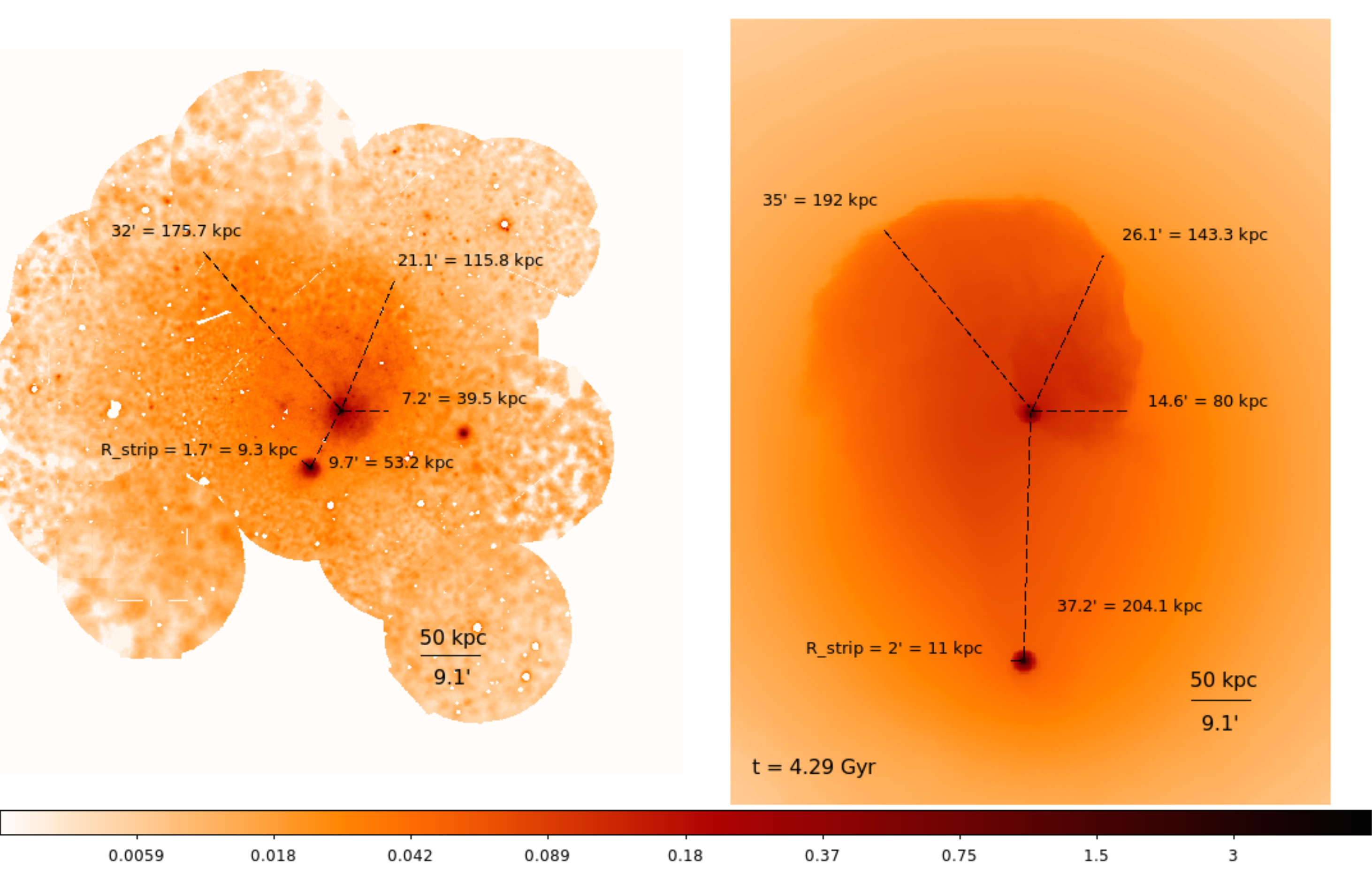}
\caption{Comparing simulation to observation. Top: An XMM image of Fornax in the 0.5 - 2.0 keV energy band alongside a mock X-ray image of the V2 simulation. This simulation timestep is chosen to provide the best match in projected distance between NGC 1404 and the Fornax center. Our simulation reproduces the prominent cold front to the north of the cluster along with a comparable stripping radius for NGC 1404. However, at this moment, the sloshing is slightly too advanced, i.e., the CF radii are somewhat too large. Given the simplicity of the simulation model, this quantitative agreement is good. Bottom: Same as top, but an earlier simulation time to provide a better distance match to the northern sloshing front. This shows that a faster infall velocity is probably required to enable NGC 1404 to reach the best match point so that the distances to the sloshing fronts are a closer match to observation. Images are in units of photons/s/cm$^{2}$/deg$^{2}$. $1'$ = 5.49 kpc.}
\label{fig:v2_xmm}
\end{figure*}

To make a direct visual comparison, Figure \ref{fig:v2_xmm} presents an XMM image of Fornax alongside a mock X-ray image for the best match in the V2 simulation in the same units. We choose this particular simulation to make the comparison rather than V0 or V1 due to its sloshing features providing the best match, including both the northern and southern cold fronts. This time step is the best compromise between matching the cold front radii and the NGC 1404 position to the observation. At a slightly earlier time, the sloshing spiral would be smaller, and match the observation, while NGC 1404 would then be even a bit farther from its desired position. Fine-tuning its original infall velocity could reconcile this but brings no new insights. Given the simplicity of the model the achieved and consistent match between all the simulations and the observation is remarkable. \par

To see how sensitive the features of the simulation are to projection effects, we took the mock X-ray image of the V2 simulation and made images using various lines of sight (LOS). The results of this are presented in Figure \ref{fig:v2_xray}. The x-y image plane as shown corresponds to the orbital plane and we vary the projection along the z axis, either perpendicular to the image plane or rotated about the horizontal or vertical axis. The overall sloshing features are independent of the LOS for a range of LOS rotated horizontally or vertically by 45$^{\circ}$ and 76$^{\circ}$. The most notable difference is that the more you rotate the view, the less visible are the KHI's along the prominent cold front to the north. Regardless of the LOS, NGC 1404 and the center of Fornax show no real change in brightness or shape.

\subsection{Predictions}
So far we have discussed the comparison of the simulations to the known features in the observation. Here we predict some features that have not yet been observed, but should exist if the proposed merger scenario is correct.

\begin{figure}[h!]
\centering
\includegraphics[scale=0.35]{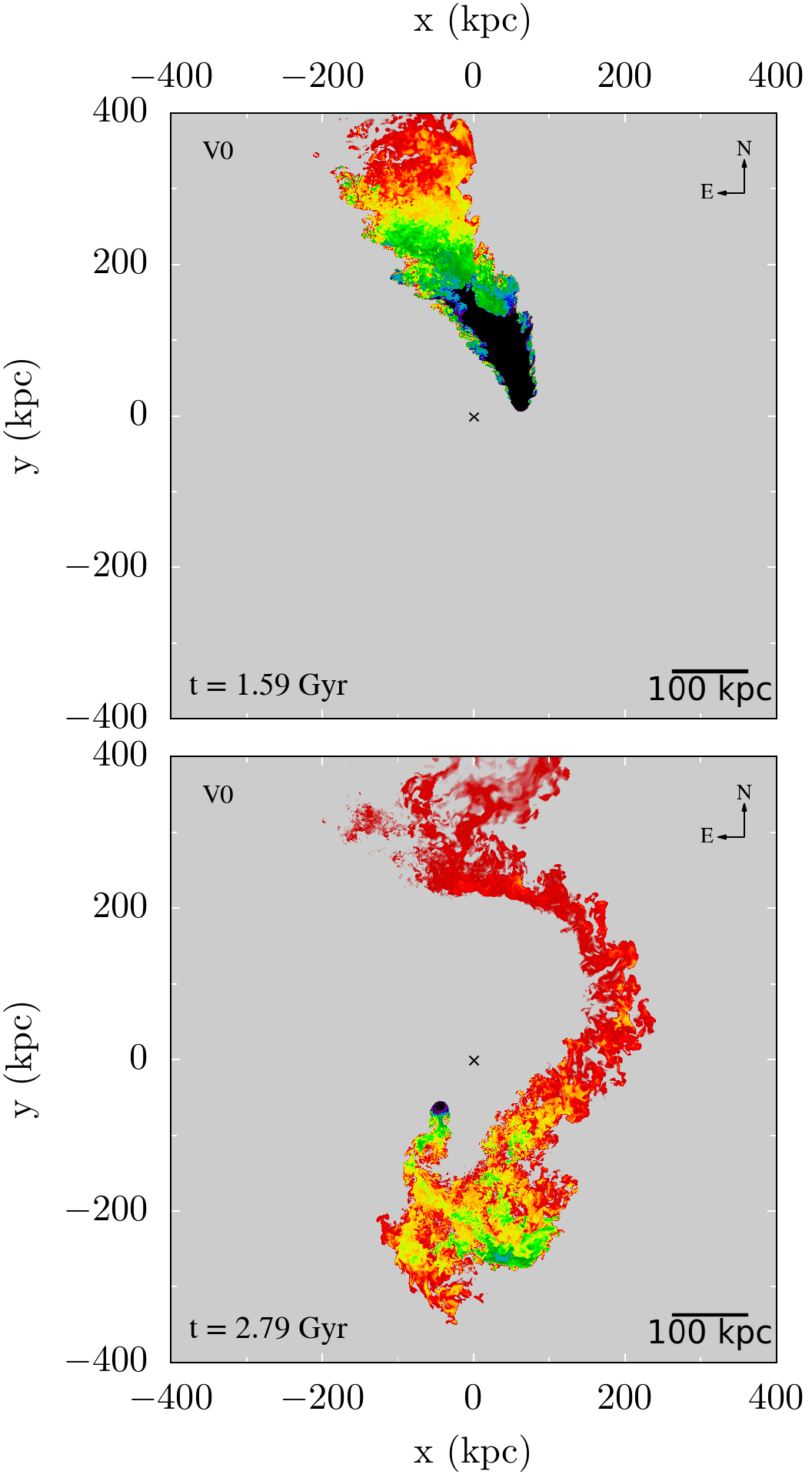}
\includegraphics[scale=0.35]{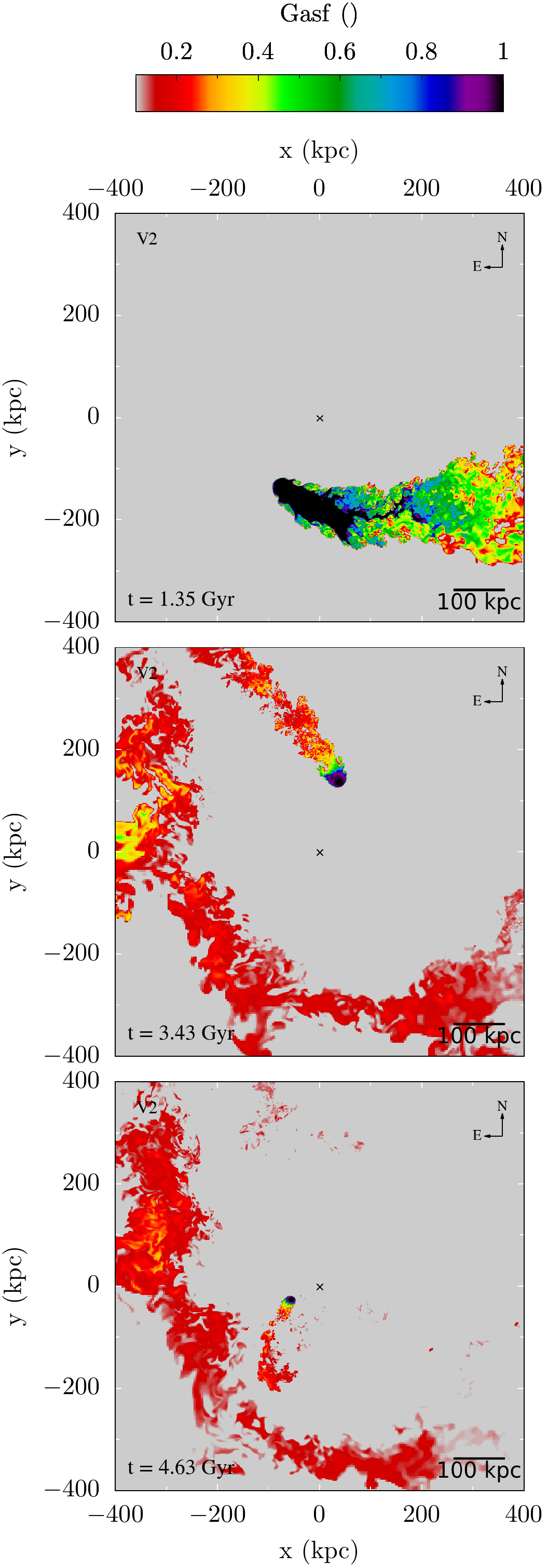}
\caption{Left: Comparison between the galactic gas fraction slices in the orbital plane for the first and second infall for the V0 simulation. Right: Same as right but also the third infall for the V2 simulation. On the first infall into the cluster, the first 100 kpc of the tail is the unmixed remnant tail, just the pushed-back atmosphere of NGC 1404 (compare Figure 4 in \citealp{Roediger2015c}). It is not until the second or third infall that the atmosphere of NGC 1404 is truly stripped. The cluster center is marked by "x". gasf = 1.0 means 100\% galactic gas.}
\label{fig:gasfcomp}
\end{figure}

\subsubsection{Detached Bow Shock Prediction}
Evident in all of the simulations is the robust feature of a detached bow shock south of NGC 1404 as indicated in Figure \ref{fig:wholeevo}. This shock is a remnant from the galaxy's previous encounter with Fornax. As the galaxy moved through pericenter, this shock led the galaxy. But, as the galaxy moves out of the cluster reaching the southern apocenter and then begins to fall back towards the center for the next infall, this shock becomes "detached" and continues to propagate outwards, while the galaxy turns back towards the Fornax center. This is a significant characteristic of the merger history and could be seen in other merging clusters such as NGC 4839 in the Coma cluster \citep{Neumann2003}. The distance of this detached bow shock in Fornax from the galaxy varies in each of the simulations between 450 - 750 kpc, depending on the angular momentum of NGC 1404. However, in each case the strength of the shock remains approximately the same. We estimate the shock Mach number between 1.3 and 1.5. At this strength, we note that this feature could potentially be observed and would therefore support our scenario of a second or third infall for NGC 1404. Also the fact that the distance of the shock appears to be dependent on the angular momentum gives us an ability to disentangle the history from observation.

\subsubsection{Wake Region + Turbulence}
This single merger between NGC 1404 and Fornax leaves a trail of destruction throughout the cluster and importantly we can identify regions where NGC 1404 did not stir any turbulence (other galaxies may have). It is important to note that a caveat in our setup is that our simulations begin as an idealized system without initial turbulence in the cluster gas, which is not the most realistic setup to use, but is adequate for the aims of this paper. From Figures \ref{fig:wholetempevo} and \ref{fig:metf} we can identify that none of the simulations predict enhanced turbulence outside the northern cold front, towards the north and north west. There is robust apocenter turbulence south of the galaxy in all cases which is embedded in a cooler region. Further the most recent wake of NGC 1404 is in the same location where we have an enhanced turbulent region along with a slight enhancement in metallicity. If we believe the slow first infall case is unlikely, then there should be old turbulence driven by an earlier infall of NGC 1404 to the south west of the cluster center, this is clearly evident in the V1 and V2 simulations. An overriding thought here is that this simulation offers an opportunity to study turbulence in clusters as we have predicted regions where we believe it should or should not have been stirred substantially.

\subsubsection{Stolen Atmosphere}
\label{sec:sa}
A particularly interesting result from our simulations is the observed hot ICM "halo" surrounding the galaxy at the second and third infall stage as shown in Figure \ref{fig:temp_xray} (right column). It is clear that the surrounding ICM is heated up to temperatures much higher than the original ICM temperature at this cluster-centric radius and this effect is slightly stronger on the second infall compared to the third infall. The hot halo seen here is essentially a "stolen atmosphere" i.e. Fornax ICM drawn into the NGC 1404 potential by gravity. The heating occurs as the cluster gas is compressed by the potential of the galaxy as it traverses the ICM. This enhancement is most noticeable on the second infall as the acceleration due to the potential of the galaxy is much greater due to its smaller gas radius, resulting in the accretion of more Fornax gas. From an observational standpoint, this stolen atmosphere effect has been seen in the dark subcluster of A520 which is undergoing a cluster merger (\citealp{H.S.Wang2016}). Furthermore, the ICM compression in the "stolen atmosphere" could lead to an over-estimate of the galaxy velocity from the standard stagnation point method. An example of this effect could possibly be seen with NGC 4472 which is falling into the Virgo cluster, as \citet{Kraft2011a} calculate very high velocities for the galaxy. We will study this effect in detail in Fish et al. (in prep).

\section{Discussion}
\label{sec:discussion}

\subsection{Merging History}
Attempting to decipher a Gyr-long merging history from a single snapshot in time may seem like an adventurous undertaking, however by simply applying the governing physics - gravity and hydrodynamics, combined with the available constraints on the cluster and galaxy gas distribution as well as the gravitational potentials, leads to a consistent recent merger history of Fornax and infall history of NGC 1404. The merger scenario proposed above explains both the stripping state of NGC 1404 and the overall sloshing features of Fornax. Our tailored merger simulations match the features of Fornax and NGC 1404 qualitatively and also almost quantitatively. We derive slightly different merger ages from the position of NGC 1404 and the positions of the sloshing cold fronts. However, this simply reflects the uncertainty in the exact infall velocity as well as the outer potential of Fornax. \par

\begin{figure}[h!]
\centering
\includegraphics[scale=0.58]{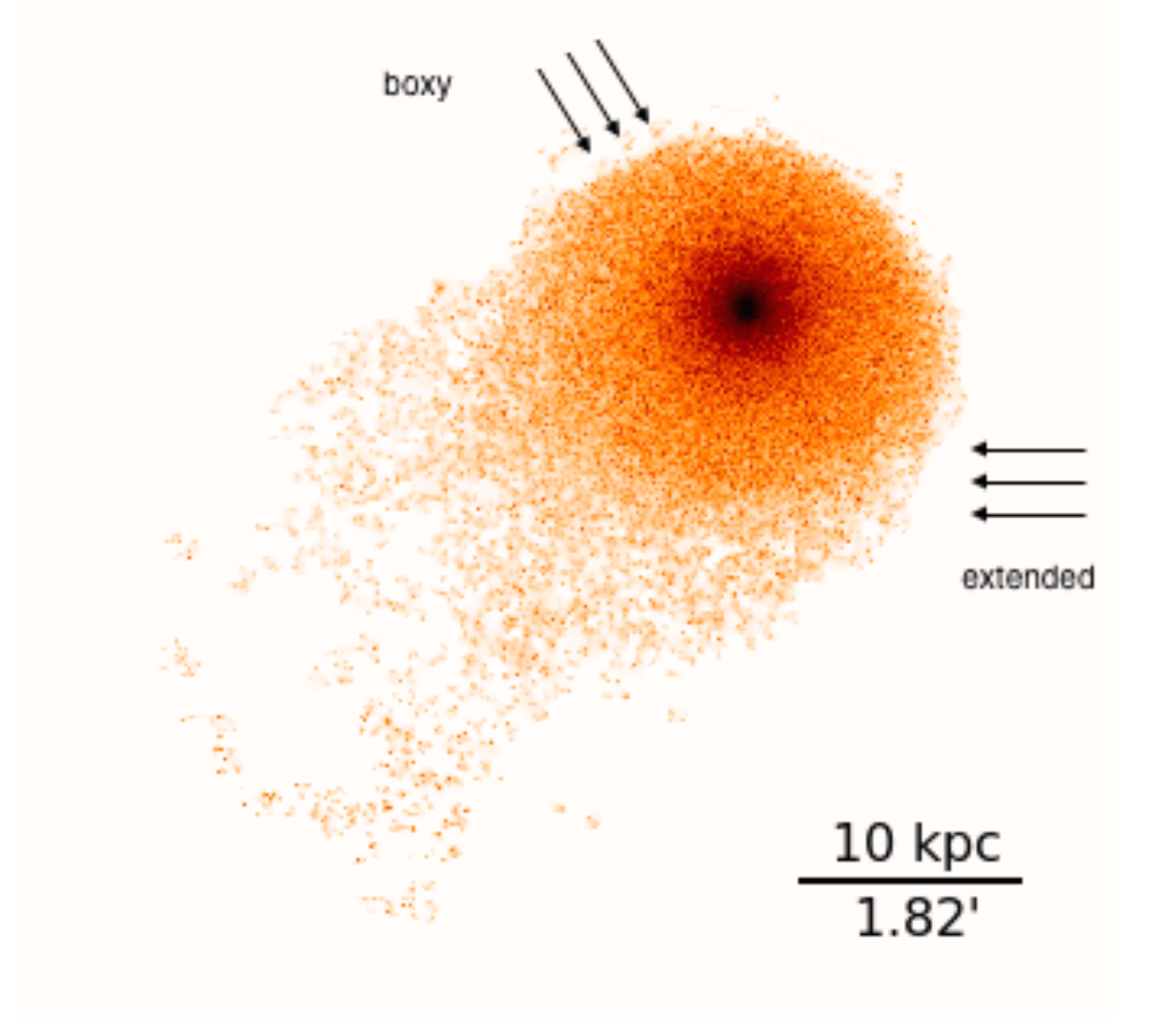}
\includegraphics[scale=0.3]{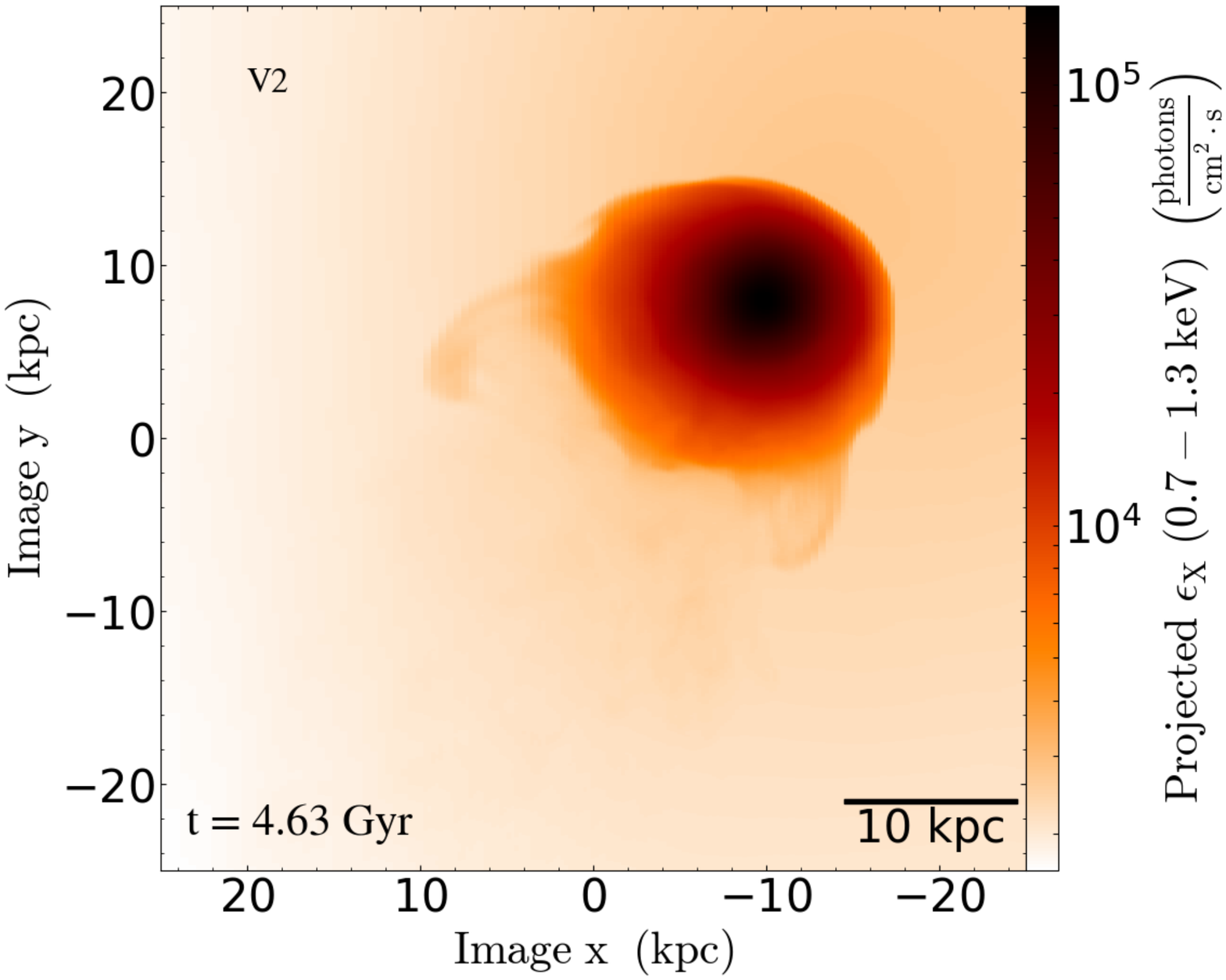}
\caption{Comparing the simulation to observation. Top: Chandra image of NGC 1404 in the  0.7 - 1.3 keV energy band taken from \citet{Su2017e}. Bottom: X-ray photon emissivity projection in the same energy band for the best match of the V2 simulation. The simulation reproduces a comparable stripping radius for NGC 1404 as well as the faint gaseous tail.}
\label{fig:v2galxrayzoom}
\end{figure}

We can rule out that NGC 1404 is still on its very first infall, unless, for an unknown reason, it already had a strongly truncated gas atmosphere prior to infall. Any extended atmosphere, as typical for isolated ellipticals, could not be stripped on first infall, but would trail the galaxy as a bright, long, cool tail, easily detectable. Further, NGC 1404 would be supersonic and a bow shock would be evident in data from Chandra and this is not the case. Figure \ref{fig:gasfcomp} illustrates that this first infall tail is indeed unmixed galactic gas. This result is consistent with the simulations of \citet{Roediger2015c} for M89 in Virgo, which also showed that the near tail of M89 can be readily explained as a remnant tail.

\begin{figure}[h!]
\centering
\includegraphics[scale=0.56]{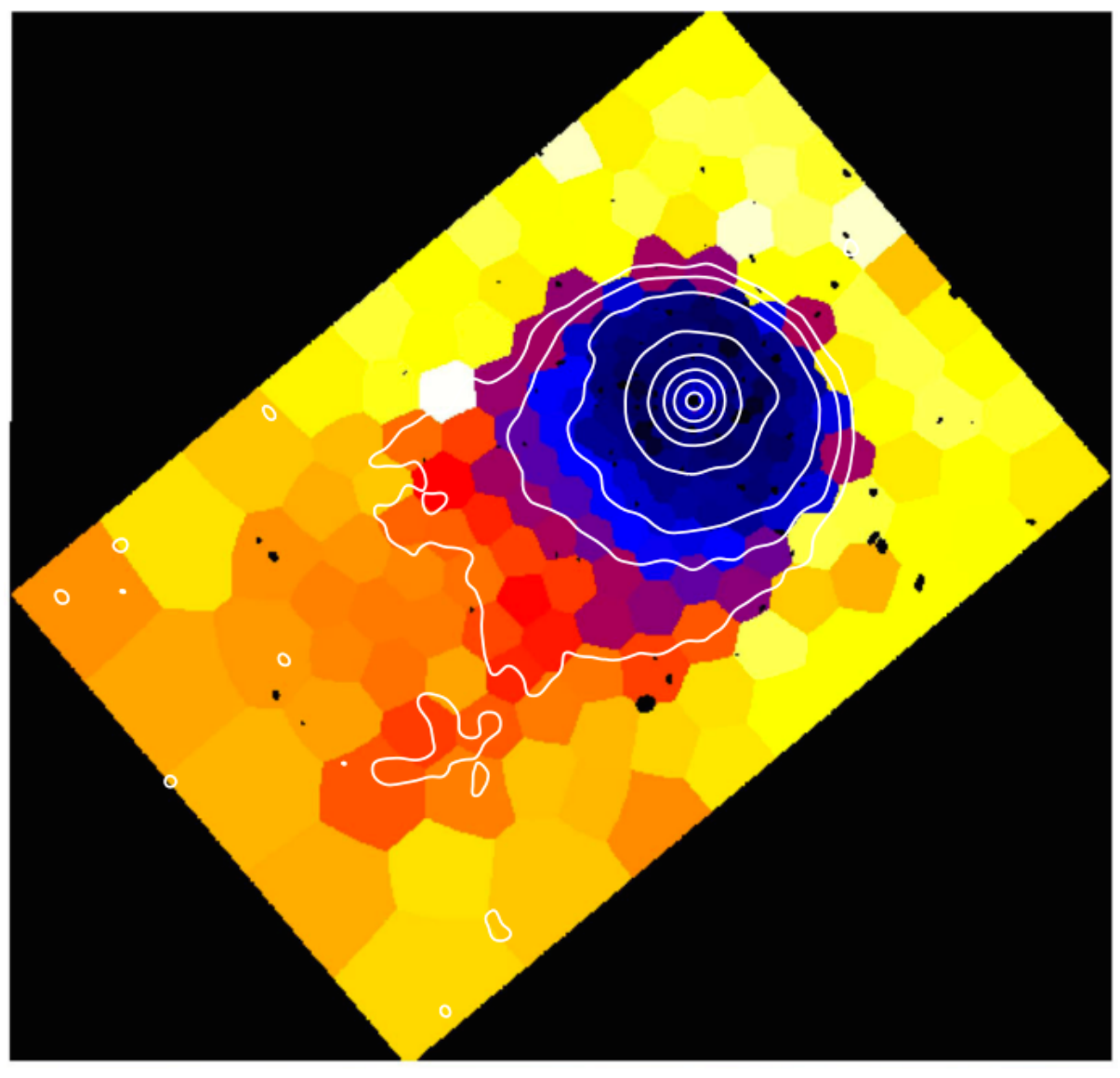}
\includegraphics[scale=0.3]{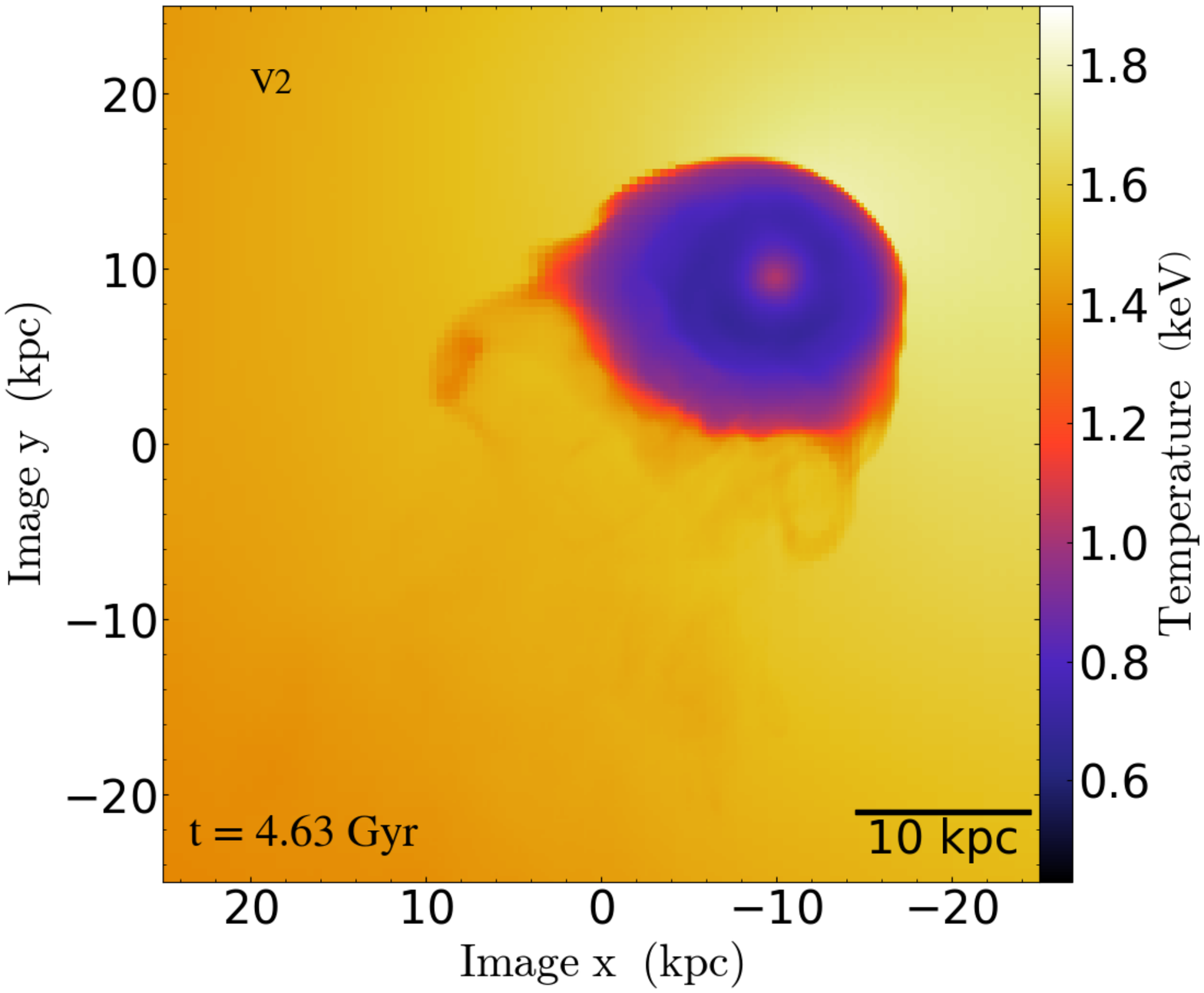}
\caption{Comparing the simulation to observation. Top: Chandra derived temperature map of NGC 1404 taken from \citet{Su2017e}. Bottom: Projected temperature map weighted by X-ray emissivity for the best match of the V2 simulation. The simulation reproduces the cooler inner atmosphere of NGC 1404. A short, cool gas tail is also just about evident in the simulation. Both have units of keV and the same colour scale.}
\label{fig:v2galtempxrayzoom}
\end{figure}

\subsection{Enrichment Processes}
Our simulations show that the interstellar medium from a single galaxy is distributed widely through the host cluster (Figure \ref{fig:metf}). \citet{Biffi2018a} conducted a study into tracing back the spatial origin of metals in the outskirts of simulated galaxy clusters. They found that in situ enrichment of the ICM in the outskirts is not a major contribution in the metallicity content of present day clusters, but that merging substructures play a more influential role in the enrichment. As demonstrated in our simulations, the NGC 1404 merger with NGC 1399, redistributes metal-rich gas that was originally in the centers of NGC 1399 and NGC 1404. Gas from NGC 1404 is distributed along its tail, although stripping of the outer, metal-poor atmosphere does not lead to enrichment of the Fornax ICM. Metal-rich gas from NGC 1399 is redistributed to \sim 200 kpc by sloshing. Thus, the entire merger history of a cluster is imprinted in the distribution of temperature and metallicity. If the notion that merging substructures play a more influential role in the enrichment of clusters is true, then the ability to access this information with XARM and the Athena XIFU in the years to come could transform our understanding of the formation of clusters.

\subsection{Importance of Dynamical Friction} 
Originally we experimented with pure hydrodynamic simulations using rigid gravitational potentials. With this method, the cluster and galaxy are not self gravitating, but instead are modelled by the sum of their individual rigid gravitational potentials that evolve on a mutual orbit. Initially, this orbit was a test particle through the more massive cluster i.e. dynamical friction was absent. This method is technically easier to deal with and reduces the computational load, but has the drawback of lacking the important effects of tidal forces and dynamical friction. Qualitatively, the simulation characteristics remain largely the same between the two physical models in the sense that we see the same gas stripping physics which produce the same features of instabilities and wakes.

 However, the lack of dynamical friction acting on the galaxy allows the galaxy to travel further out of the cluster after the first pericenter passage and consequently it lingers for too long around the apocenter, thus resulting in a greatly increased merger timescale. This also has an impact on the strength of the ram pressure the galaxy feels as this is dependent on the square of the velocity, and therefore impacts how much gas is stripped from the galaxy. Comparing the two methods and using the same simulation design as in V0, we found that including dynamical friction reduces the orbiting radius and the merging time scale by 60\% and 55\%, respectively.

\subsection{NGC 1404 as a Test Case for ICM Plasma Physics} 
Both merger history and ICM plasma physics could affect ICM properties (tail, wake, cold fronts etc). Our work demonstrates that before using observations to probe ICM plasma physics, we need to understand the merger history. It is worth emphasizing that we have reproduced the short tail, boxy front, and the temperature gradient in the tail. Our simulations are inviscid (assuming numerical viscosity is small). Therefore, the Fornax ICM must have a low viscosity.
As we have determined the dynamics of the Fornax ICM, we can confidently use this cluster as a test case for ICM properties such as its effective viscosity and thermal conductivity. Viscosity or aligned magnetic fields could prevent mixing of cold and warmer ICM gas in two locations: along the sloshing cold fronts around the Fornax center and in the wake of NGC 1404. The near wake would be the best target as it is easiest to observe. Suppressing mixing in the wake should lead to a brighter, cooler tail for NGC 1404 even if its atmosphere is already pre-truncated due to the previous cluster passages. We will investigate this question in a forthcoming paper.

\subsection{Globular Cluster Debate}
An interesting point to note is that the second/third infall could offer an explanation to the globular cluster content debate of NGC 1404 and NGC 1399. It has been suggested through measurements of their specific frequency that NGC 1404 has lost some of its globular clusters to NGC 1399 (\citealp{Forbes1998a}, \citealp{Bekki2003a}) due to a possible interaction via tidal stripping. Our proposed scenario requires that NGC 1404 passed by NGC 1399 at least once, where the loss of globular clusters could have happened. Especially if the third infall scenario is the correct case, this gives two opportunities for NGC 1404 and NGC 1399 to interact and exchange globular clusters. It should be noted here that our simulations differ greatly from that in \citet{Bekki2003a} in terms of the galaxy orbit. In their simulation, the orbit radius does not extend past 60 kpc compared to reaching between 280-450 kpc in our simulations. Our simulation is more advanced than \citet{Bekki2003a} and we have direct access to tidal forces. If the particles are traceable and can be considered as representative of GCs, we can quantify the tidal stripping process and compare to the observed and kinematical properties of the GCs (future work).
 
\section{Summary}
\label{sec:summary}
We used simple hydro+Nbody tailored simulations to model the infall of NGC 1404 into the Fornax Cluster in an attempt to analyse its recent evolutionary history. Our simulations varied the infall velocity between almost zero at \sim  virial radius to an initial approach velocity of Mach 1 at the virial radius. Furthermore, we varied the impact parameter of the merger, independent of these choices, and our results reveal a consistent picture over the last 1.1 - 1.3 Gyrs. Our results are as follows: \\
\begin{itemize}
 \item We can firmly exclude that NGC 1404 is on its first infall into the cluster. It is either on a second or (more likely) third infall due to the position of the sloshing fronts in Fornax. 
 \item NGC 1404 came from the E or NE and passed by NGC 1399 about 1.1 - 1.3 Gyrs ago and is now at the point of its next encounter.
 \item This merging history can explain the sloshing patterns observed in Fornax and the truncated atmosphere and stripping radius of NGC 1404. Furthermore, this scenario also reproduces the observed temperature and abundance asymmetries observed by \citet{Murakami2011c}. 
\end{itemize}
In our simulations, the exact position of NGC 1404 is not matched and requires fine tuning. The features produced in our simulations are robust that making a more accurate match to the galaxy's position would not change the overall merger scenario sufficiently. Independent of the exact history of the merger, several features remain robust and therefore we can make the following predictions:

\begin{itemize}
  \item A detached bow shock at a distance between 450-750 kpc south of NGC 1404, a remnant of the galaxy's previous infall, should exist with an estimated Mach number between 1.3 and 1.5.
  \item The most recent wake of NGC 1404 lies S of the galaxy with enhanced turbulence and a slight enhancement in metallicity.
  \item  If we believe a slow first infall is unlikely, then there should be old turbulence from a previous infall to the SW of NGC 1404.
  \item  None of our simulations predicts enhanced turbulence outside of the cold front N, NW of the cluster center. That is to say, NGC 1404 did not stir turbulence in this region.
\end{itemize}

\acknowledgments
The software used in this work was in part developed by the DOE NNSA-ASC OASCR Flash Center at the University of Chicago. \software{FLASH (\citealp{Fryxell2000})} \\

ER acknowledges the support of STFC, through the University of Hull's Consolidated Grant ST/R000840/1 and access to viper, the University of Hull High Performance Computing Facility.

\bibliography{template}

\end{document}